%% file: main.tex
\documentclass[a4paper]{article}
\usepackage[T1]{fontenc}
\usepackage{amsmath}
\usepackage{amssymb}
\usepackage{latexsym}
\usepackage{bussproofs}
\usepackage{cmll}

\usepackage{version}

\usepackage[dvips]{color}
\usepackage{epsfig}

\usepackage{array}
\usepackage{alltt}
\usepackage{theorem}
\usepackage{array}
\usepackage[all]{xy}
\usepackage{graphicx}

\usepackage{tikz}
\usetikzlibrary{matrix}

\usepackage{hyperref}

\input{local}

\makeatletter
\def\@setcopyright{}
\def\serieslogo@{}
\makeatother

\title{Full abstraction for probabilistic PCF}
\author{Thomas Ehrhard, Michele Pagani and Christine Tasson\\
  CNRS, IRIF,
  Univ Paris Diderot, Sorbonne Paris Cit\'e\\
  F-75205 Paris, France
 }

\input{notation}

\begin{document}
\allowdisplaybreaks

\maketitle
\begin{abstract}
  We present a probabilistic version of PCF, a well-known simply typed
  universal functional language. The type hierarchy is based on a single ground
  type of natural numbers. Even if the language is globally call-by-name, we
  allow a call-by-value evaluation for ground type arguments in order to
  provide the language with a suitable algorithmic expressiveness. We describe
  a denotational semantics based on probabilistic coherence spaces, a model of
  classical Linear Logic developed in previous works. We prove an adequacy and
  an equational full abstraction theorem showing that equality in the model
  coincides with a natural notion of observational equivalence.
\end{abstract}

\section*{Introduction}
PCF is a paradigmatic functional programming language introduced by Dana Scott
in~1969 and further studied by many authors, see in
particular~\cite{Plotkin77}. 

The denotational semantics of PCF and of its extensions by various kinds of
effects is one of the major research topics in the semantics of programming
languages because of the relative simplicity of the language combined with its
computational expressiveness and because of its extremely clean and canonical
mathematical semantics. The development of major functional programming
languages such as \textsf{Ocaml}, \textsf{Haskell} or \textsf{F\#} benefited
from these theoretical studies. 

As far as purely functional features of these languages are considered, the
standard setting of cartesian closed categories with a distinguished object for
interpreting natural numbers and fix-point operators are sufficient. Most
considered such categories have complete partially ordered sets as objects and
all Scott continuous functions as morphisms. Extending such models with a
probabilistic effect in order to build a model of a probabilistic functional
language is a notoriously difficult problem, especially if we insist on objects
to contain a reasonably ``small'' dense subset\footnote{By this we mean that
  all objects of the sought category should contain a dense subset whose
  cardinality is less than a fixed cardinal. This is necessary in particular if
  we want the category to host models of the pure lambda-calculus.}, see in
particular~\cite{JungTix98}. More precisely, it seems very difficult to find
cartesian closed categories of continuous domains equipped with a probabilistic
powerdomain monad.

There is however another approach to the denotational semantics of
probabilistic functional languages. Initiated in~\cite{Girard04a} (based on
earlier quantitative ideas coming from~\cite{Girard88c,Girard99,Ehrhard00c}),
this theory of \emph{probabilistic coherence spaces} was further developed
in~\cite{DanosEhrhard08} where it has been shown to provide a model of
classical Linear Logic allowing also to interpret arbitrary fix-points of
types, and hence to host many models of the pure lambda-calculus. We further
studied this semantics in~\cite{EhrhardPaganiTasson11} where we proved an
adequacy theorem in a pure lambda-calculus model, and
in~\cite{EhrhardPaganiTasson14} we also proved a full abstraction theorem for a
probabilistic version of PCF, interpreted in the probabilistic coherence space
model.


The goal of the present paper is to provide a more detailed presentation of
this full abstraction result, recording also the proof of adequacy. With
respect to~\cite{EhrhardPaganiTasson14}, our new presentation provides a major
improvement concerning the syntax of the programming language under
consideration.

Indeed, in this previous work we considered a \emph{fully call-by-name} (CBN)
version of probabilistic PCF. In this language, a closed term of type $\Tnat$
(the type of natural numbers) determines a sub-probability distribution on the
natural numbers (with $n$ we associate the probability that $M$ reduces 
to the constant $\Num n$ of the language). A closed term $P$ of type
$\Timpl\Tnat\Tnat$ which receives $M$ as argument will reduce $M$ each time it
needs its value (because the language is fully CBN) and will get different
results each time unless the sub-probability distribution defined by $M$ is
concentrated on a single natural number. There are clearly cases where this is
not a desirable behavior: sometimes we need to flip a coin and to use the result
several times!

As an example, consider the problem of writing a program which
takes an array $f$ of integers of length $n$ and returns an index $i$ such that
$f(i)=0$; we want to apply a ``Las Vegas'' random algorithm consisting in
choosing $i$ randomly (with a uniform probability on $\{0,\dots,n-1\}$)
repeatedly until we find an $i$ such that $f(i)=0$. This is implemented by
means of a \textsf{while} loop (or, more precisely, of a recursively defined
function since we are in a functional setting) where at each step we choose $i$
randomly, test the value of $f(i)$ (first use of $i$) and return $i$ (second
use) if $f(i)=0$. It is intuitively clear that this basic algorithm cannot be
implemented with the usual conditional of PCF (it might be interesting and
challenging to prove it).

To be able to write such an algorithm, we need to modify PCF a bit, allowing to
use ground terms in a call-by-value (CBV) fashion (to simplify the presentation
we use $\Tnat$ as single ground type).

Our choice has been to modify the conditional construct.  The usual
conditional construct $\If MPQ$ of PCF is operationally interpreted as follows:
one first reduces $M$ until one gets an integer $n$ (or, more precisely, the
corresponding term $\Num n$). If $n=0$, one evaluates $P$ and otherwise, one
evaluates $Q$, in the current context of course. Again, the trouble is that, in
the second case, the value obtained for $M$, namely $\Num n$, is lost, whereas
$Q$ might need it. This problem can be easily solved by using $M$ within $Q$
each time this value is needed. Although clearly inefficient, this solution is
perfectly correct in the usual deterministic version of $\PCF$. It is
absolutely inadequate in our probabilistic setting since $M$ should be
considered as a \emph{probabilistic process} whose reduction, or execution,
will produce integer values with a sub-probability distribution depending on
it. There is no reason for $M$ to produce, within $Q$, the same result $\Num n$
that it reduced to during its first evaluation.

For these reasons, when $M$ reduces to $\Num{n+1}$, our conditional
construction $\Ifv{M}{P}{z}{Q}$ allows to feed $Q$ with $\Num{n}$ through the
variable $z$ (this has the positive side effect of making the predecessor
function definable); in other words we have the reduction rules
\begin{center}
  \AxiomC{}
  \UnaryInfC{$\Ifv{\Num{0}}{P}{z}{Q}\Rel\rightarrow P$}
  \DisplayProof
  \quad
  \AxiomC{}
  \UnaryInfC{$\Ifv{\Num{n+1}}{P}{z}{Q}\Rel\rightarrow\Subst Q{\Num n}z$}
  \DisplayProof
\end{center}

\begin{center}
  \AxiomC{$M\Rel\rightarrow M'$}
  \UnaryInfC{$\Ifv{M}{P}{z}{Q}\Rel\rightarrow\Ifv{M'}{P}{z}{Q}$}
  \DisplayProof
\end{center}
This means that our conditional construct allows to use a CBV reduction
strategy, limited to the ground type of natural numbers. 

From the point of view of Linear Logic and of its denotational models, this
feature is completely justified by the fact that the object interpreting the
type of natural numbers has a canonical structure of coalgebra for the $\oc$
exponential functor. Intuitively, this means that \emph{evaluated natural
  numbers} can be freely discarded and duplicated. Pushing this idea further
leads to consider a calculus~\cite{Ehrhard15b} close to Levy's
Call-By-Push-Value~\cite{LevyP06} whose probabilistic version will be
considered in a forthcoming paper.

\paragraph*{Contents.}
We present the syntax of Probabilistic PCF ($\PPCF$) and its weak-reduction
relation, that we formalize as an infinite dimensional stochastic matrix
(indexed by $\PPCF$ terms). Based on this operational semantics, we define a
notion of observational equivalence. 
Two terms of type $\sigma$ in a typing context $\Gamma$ are equivalent if, for
any context $\Thole C\Gamma\sigma$ of type $\Tnat$ in context $\Gamma$ (with
holes of type $\sigma)$, the probability that $\Thsubst CM$ reduces to $\Num 0$
(say) is equal to the probability that $\Thsubst {C}{M'}$ reduces to $\Num 0$.

Then we give various examples of programs written in this language, some of
them will be essential in the proof of the Full Abstraction Theorem. In
particular we implement the above mentioned simple Las Vegas algorithm.

Next, we introduce the model of Probabilistic Coherence Spaces (PCS), presented
as a model of classical Linear Logic. We describe the interpretation of $\PPCF$
terms, presenting the semantics of terms as functions (this is possible because
the Kleisli category of the $\oc$-comonad of this model is well-pointed). We
prove an Adequacy Theorem which states that, for any closed term $M$ of ground
type $\Tnat$ and any $n\in\Nat$, the probability that $M$ reduces to $\Num n$ is
equal to the probability of $n$ in the sub-probability distribution on $\Nat$
which is the semantics of $M$ in the PCS model. This implies that any two
closed terms of type $\sigma$ which have the same interpretation in PCS are
observationally equivalent.

Last we prove the converse implication showing that PCS is a Fully Abstract
model of $\PPCF$. The proof uses strongly the fact that, in our model,
morphisms are analytic functions (with real non-negative coefficients) and that
the coefficients of the entire series of two such functions are the same if the
functions coincide on an open subset of their domain.  Section~\ref{sec:fullabs}
is devoted to this theorem and to its detailed proof; it starts with a more
accurate description of our proof method.


\section{Probabilistic PCF}

There is only one ground type $\Tnat$, types are defined by
\begin{align*}
  \sigma,\tau,\dots{} \Bnfeq \Tnat \Bnfor\Impl\sigma\tau
\end{align*}

The terms of $\PCFP$ are defined as follows:
\begin{align*}
  M,N,\dots{} \Bnfeq \Num n
  &\Bnfor x \Bnfor \Succ M \Bnfor \Ifv MPzR \Bnfor \Abst
  x\sigma M \Bnfor \App MN\\
  &\Bnfor \Dice p \Bnfor \Fix M
\end{align*}
where $n\in\Nat$, $p\in\Rseg 01\cap\Rational$ is a probability and $x$,
$y$\dots{} are variables.

A typing context is a sequence $\Gamma=(x_1:\sigma_1,\dots,x_n:\sigma_n)$ where
the $x_i$'s are pairwise distinct variables. A typing judgment is an expression
$\Tseq\Gamma M\sigma$ where $\Gamma$ is a typing context, $M$ is a term and
$\sigma$ is a type. The typing rules are as follows:
\begin{center}
  \AxiomC{}
  \UnaryInfC{$\Tseq\Gamma{\Num n}\Tnat$}
  \DisplayProof
  \quad
  \AxiomC{}
  \UnaryInfC{$\Tseq{\Gamma,x:\sigma}{x}\sigma$}
  \DisplayProof
  \quad
  \AxiomC{$\Tseq\Gamma M\Tnat$}
  \UnaryInfC{$\Tseq\Gamma{\Succ M}\Tnat$}
  \DisplayProof
\end{center}
\begin{center}
  \AxiomC{$\Tseq\Gamma M\Tnat$}
  \AxiomC{$\Tseq\Gamma P\sigma$}
  \AxiomC{$\Tseq{\Gamma,z:\Tnat}{R}\sigma$}
  \TrinaryInfC{$\Tseq\Gamma{\Ifv MPzR}\sigma$}
  \DisplayProof
\end{center}
\begin{center}
  \AxiomC{$\Tseq{\Gamma,x:\sigma}M\tau$}
  \UnaryInfC{$\Tseq\Gamma{\Abst x\sigma M}{\Timpl\sigma\tau}$}
  \DisplayProof
  \quad
  \AxiomC{$\Tseq\Gamma M{\Timpl\sigma\tau}$}
  \AxiomC{$\Tseq\Gamma N\sigma$}
  \BinaryInfC{$\Tseq\Gamma{\App MN}\tau$}
  \DisplayProof
  \quad
  \AxiomC{$\Tseq\Gamma M{\Timpl\sigma\sigma}$}
  \UnaryInfC{$\Tseq\Gamma{\Fix M}\sigma$}
  \DisplayProof
\end{center}
\begin{center}
  \AxiomC{$p\in[0,1]\cap\Rational$}
  \UnaryInfC{$\Tseq\Gamma{\Dice p}\Tnat$}
  \DisplayProof
\end{center}

\begin{proposition}\label{prop:syntax-typing}
  Let $M$ be a term and $\Gamma$ be a typing context. There is at most one type
  $\sigma$ such that $\Tseq\Gamma M\sigma$.
\end{proposition}
The proof is a simple inspection of the typing rules.

Given terms $M$ and $N$ and given a variable $x$, we use $\Subst MNx$ for the
term $M$ where $x$ is substituted with $N$. 

\begin{lemma}
If $\Tseq{\Gamma,x:\sigma}{M}{\tau}$ and $\Tseq{\Gamma}{N}{\sigma}$, then
$\Tseq{\Gamma}{\Subst MNx}{\tau}$.  
\end{lemma}
The proof us a simple induction on the structure of $M$.

\subsection{Reduction rules}\label{sec:red-rules}
Given two terms $M$, $M'$ and a real number $p\in[0,1]$, we define
$M\Rel{\Redone p} M'$, meaning that $M$ reduces in one step to $M'$ with
probability $p$, by the following deduction system.


We define first a deterministic reduction relation $\Redoned$ as follows.
\begin{center}
  \AxiomC{}
  \UnaryInfC{$\App{\Abst x\sigma M}{N}\Rel\Redoned\Subst MNx$}
  \DisplayProof
  \quad
  \AxiomC{}
  \UnaryInfC{$\Fix M\Rel\Redoned\App M{\Fix M}$}
  \DisplayProof
\end{center}
\begin{center}
  \AxiomC{}
  \UnaryInfC{$\Succ{\Num n}\Rel{\Redoned}\Num{n+1}$}
  \DisplayProof
  \quad
  \AxiomC{}
  \UnaryInfC{$\Ifv{\Num 0}PzR\Rel{\Redoned}P$}
  \DisplayProof
  \quad
  \AxiomC{}
  \UnaryInfC{$\Ifv{\Num{n+1}}PzR\Rel{\Redoned}\Subst R{\Num n}z$}
  \DisplayProof
\end{center}

Then we define the probabilistic reduction by the following rules.
\begin{center}
  \AxiomC{$M\Rel{\Redoned}M'$}
  \UnaryInfC{$M\Rel{\Redone 1}M'$}
  \DisplayProof
  \quad
  \AxiomC{}
  \UnaryInfC{$\Dice p\Rel{\Redone p}\Num 0$}
  \DisplayProof
  \quad
  \AxiomC{}
  \UnaryInfC{$\Dice p\Rel{\Redone{1-p}}\Num 1$}
  \DisplayProof
\end{center}
\begin{center}
  \AxiomC{$M\Rel{\Redone p}M'$}
  \UnaryInfC{$\App MN\Rel{\Redone p}\App{M'}N$}
  \DisplayProof
  \quad
  \AxiomC{$M\Rel{\Redone p}M'$}
  \UnaryInfC{$\Succ M\Rel{\Redone p}\Succ{M'}$}
  \DisplayProof
\end{center}
\begin{center}
  \AxiomC{$M\Rel{\Redone p}M'$}
  \UnaryInfC{$\Ifv MPzR\Rel{\Redone p}\Ifv{M'}PzR$}
  \DisplayProof
\end{center}
This reduction can be called \emph{weak-head reduction} (or simply weak
reduction) since it always reduces the leftmost outermost redex and never
reduces redexes under abstractions.  We say that $M$ is \emph{weak-normal} if
there is no reduction $M\Rel{\Redone p}M'$.



\subsection{Observational equivalence}\label{sec:obseq}
Using this simple probabilistic reduction relation, we want now to define a
notion of observational equivalence. For this purpose, we need first to
describe as simply as possible the ``transitive closure'' of the probabilistic
reduction relation defined in Section~\ref{sec:red-rules}. We represent this
relation as a matrix $\Redmats$ indexed by terms, the number $\Redmats_{M,M'}$
being the probability of $M$ to reduce to $M'$ in one step. We add also that
$\Redmats_{M,M}=1$ if $M$ is weak-normal for the weak-reduction (that is, no
reduction is possible from $M$); in all other cases we have
$\Redmats_{M,M'}=0$.  In other words, we consider the reduction as a discrete
time Markov chain whose states are terms, stationary states are weak-normal
terms and whose associated stochastic matrix is $\Redmats$. Saying that
$\Redmats$ is stochastic means that the coefficients of $\Redmats$ belong to
$\Rseg 01$ and that, for any given term $M$, one has
$\sum_{M'}\Redmats_{M,M'}=1$ (actually there are at most two terms $M'$ such
that $\Redmats_{M,M'}\not=0$). Then if $M'$ is normal, $\Redmats^k_{M,M'}$
(where $\Redmats^k=\overbrace{\Redmats\cdots\Redmats}^k$ is the $k$th power of
$\Redmats$ for the matricial product)
represents the probability of $M$ to reduce to $M'$ in \emph{at most $k$ steps}
and we obtain the probability of $M$ to reduce to $M'$ by taking the lub of
these numbers; to obtain this effect our assumption that $M'$ is a stationary
state is crucial. We explain this in more details now, considering first the
case of a general stochastic matrix $S$ indexed by a countable set $I$ of
states.

\paragraph{Probability of convergence to a stationary state.}
Let $I$ be a countable set and let $S\in[0,1]^{I\times I}$ to be understood as
a matrix with $I$-indexed rows and columns. One says that $S$ is stochastic if
$\forall i\in I\ \sum_{j\in I}S_{i,j}=1$. Given two such matrices $S$ and $T$,
their product $ST$ is given by $\forall(i,j)\in I^2\ (ST)_{i,j}=\sum_{k\in
  I}S_{i,k}T_{k,j}$ and is also a stochastic matrix.

Let $I_1^S$ be the set of stationary states, $I_1^S=\{i\in I\St S_{i,i}=1\}$
(so that if $i\in I^S_1$ and $S_{i,j}\not=0$ then $i=j$). Let $(i,j)\in I\times
I_1^S$. Then the $n$-indexed sequence $(S^n)_{i,j}\in[0,1]$ is
monotone. Indeed, for all $n$ we have
\begin{align*}
  (S^{n+1})_{i,j}
  &= \sum_{k\in I} (S^n)_{i,k}S_{k,j}\\
  &\geq (S^n)_{i,j}S_{j,j}=(S^n)_{i,j}
\end{align*}
So we can define a matrix $S^\infty\in[0,1]^{I\times I}$ as follows
\begin{align*}
  (S^\infty)_{i,j}=
  \begin{cases}
    \sup_{n\in\Nat}(S^n)_{i,j} & \text{if }(i,j)\in I\times I_1^S\\
    0 & \text{otherwise.}
  \end{cases}
\end{align*}
The matrix $S^\infty$ is a sub-stochastic matrix because, given $i\in I$
\begin{align*}
  \sum_{j\in I}(S^\infty)_{i,j}
  &= \sum_{j\in I_1^S}\sup_{n\in\Nat}(S^n)_{i,j}\\
  &= \sup_{n\in\Nat}\sum_{j\in I_1^S}(S^n)_{i,j}
  \quad\text{by the monotone convergence theorem}\\
  &\leq \sup_{n\in\Nat}\sum_{j\in I}(S^n)_{i,j}= 1
\end{align*}
Let $i,j\in I$. A \emph{path} from $i$ to $j$ is a sequence $w=(i_1,\dots,i_k)$
of elements of $I$ (with $k\geq 1$) such that $i_1=i$, $i_k=j$ and
$i_k\not=i_l$ for all $l\in\{1,\dots,k-1\}$. The \emph{weight} of $w$ is
$\Probw w=\prod_{l=1}^{k-1}S_{i_l,i_{l+1}}$. The \emph{length} of $w$ is
$k-1$. We use $\Spath ij$ to denote the set of all paths from $i$ to $j$.
\begin{lemma}\label{lemma:stochinf-paths}
  Let $(i,j)\in I\times I_1^S$. One has
  \begin{align*}
    S^\infty_{i,j}=\sum_{w\in\Spath ij}\Probw w\,.
  \end{align*}
\end{lemma}
The proof is easy. In order to obtain this property, it is important in the
definition of paths that the last element does not occur earlier.

\paragraph{The stochastic matrix of terms.}\label{par:stoch-mat-red}
Let $\Gamma$ be a typing context and $\sigma$ be a type. Let
$\Open\Gamma\sigma$ be the set of all terms $M$ such that
$\Tseq{\Gamma}M\sigma$. In the case where $\Gamma$ is empty, and so the
elements of $\Open{\Gamma}{\sigma}$ are closed, we use $\Closed\sigma$ to
denote that set. 

Let $\Redmato\Gamma\sigma\in[0,1]^{\Open\Gamma\sigma\times\Open\Gamma\sigma}$
be the matrix (indexed by terms typable of type $\sigma$ in context $\Gamma$)
given by
\begin{equation*}
  \Redmato\Gamma\sigma_{M,M'}=
  \begin{cases}
    p & \text{if }M\Rel{\Redone p}M'\\
    1 & \text{if $M$ is weak-normal and $M'=M$}\\
    0 & \text{otherwise.}
  \end{cases}
\end{equation*}
This is a stochastic matrix. We also use the notation $\Redmat\sigma$ for the
matrix $\Redmato\Gamma\sigma$ when the typing context is empty.

When $M'$ is weak-normal, the number $p=\Redmato\Gamma\sigma^\infty_{M,M'}$ is
the probability that $M$ reduces to $M'$ after a finite number of steps by
Lemma~\ref{lemma:stochinf-paths}. We write $M\Convproba pM'$ if $M'$ is
weak-normal and $p=\Redmato\Gamma\sigma^\infty_{M,M'}$.

\paragraph{Observation contexts.}
We define a syntax for observation contexts with several typed holes, all holes
having the same type. They are defined exactly as terms, adding a new
``constant symbol'' $\Hole\Gamma\sigma$ where $\Gamma$ is a typing context and
$\sigma$ is a type, which represents a hole which can be filled with a term $M$
such that $\Tseq{\Gamma}{M}{\sigma}$. Such an observation context will be
denoted with letters $C$, $D$\dots{}, adding $\Gamma\vdash\sigma$ as
superscript for making explicit the typing judgment of the terms to be inserted
in the hole of the context. So if $C$ is an observation context with holes
$\Hole\Gamma\sigma$, this context will often be written $\Thole C\Gamma\sigma$
and the context where all holes have been filled with the term $M$ will be
denoted $\Thsubst CM$: this is just an ordinary $\PPCF$ term. Notice that, in
$\Thsubst CM$, some (possibly all) free variables of $M$ can be bound by
$\lambda$'s of $C$. For instance, if $C=\Abst
x\sigma{\Hole{x:\sigma}{\sigma}}$, then $\Thsubst Cx=\Abst x\sigma x$.

More formally, we give now the typing rules for observation contexts.
\begin{center}
  \AxiomC{}
  \UnaryInfC{$\Tseq{\Gamma,\Delta}{\Hole\Delta\tau}{\tau}$}
  \DisplayProof
\end{center}
\begin{center}
  \AxiomC{}
  \UnaryInfC{$\Tseqh\Gamma{\Num n}\Delta\tau\Tnat$}
  \DisplayProof
  \quad
  \AxiomC{}
  \UnaryInfC{$\Tseqh{\Gamma,x:\sigma}{x}\Delta\tau\sigma$}
  \DisplayProof
  \quad
  \AxiomC{$\Tseqh\Gamma C\Delta\tau\Tnat$}
  \UnaryInfC{$\Tseqh\Gamma{\Succ C}\Delta\tau\Tnat$}
  \DisplayProof
\end{center}
\begin{center}
  \AxiomC{$\Tseqh\Gamma C\Delta\tau\Tnat$}
  \AxiomC{$\Tseqh\Gamma D\Delta\tau\sigma$}
  \AxiomC{$\Tseqh{\Gamma,z:\Tnat}{E}\Delta\tau\sigma$}
  \TrinaryInfC{$\Tseqh\Gamma{\Ifv CDzE}\Delta\tau\sigma$}
  \DisplayProof
\end{center}
\begin{center}
  \AxiomC{$\Tseqh{\Gamma,x:\sigma}C\Delta\phi\tau$}
  \UnaryInfC{$\Tseqh\Gamma{(\Abst x\sigma C)}\Delta\phi{\Timpl\sigma\tau}$}
  \DisplayProof
\end{center}
\begin{center}
  \AxiomC{$\Tseqh\Gamma C\Delta\phi{\Timpl\sigma\tau}$}
  \AxiomC{$\Tseqh\Gamma D\Delta\phi\sigma$}
  \BinaryInfC{$\Tseqh\Gamma{\App CD}\Delta\phi\tau$}
  \DisplayProof
  \quad
  \AxiomC{$\Tseqh\Gamma C\Delta\tau{\Timpl\sigma\sigma}$}
  \UnaryInfC{$\Tseqh\Gamma{\Fix C}\Delta\tau\sigma$}
  \DisplayProof
\end{center}
\begin{center}
  \AxiomC{$p\in[0,1]\cap\Rational$}
  \UnaryInfC{$\Tseqh\Gamma{\Dice p}\Delta\tau\Tnat$}
  \DisplayProof
\end{center}

\begin{lemma}
  If $\Tseqh{\Gamma}{C}{\Delta}{\tau}{\sigma}$ and $\Tseq\Delta M\tau$, then
  $\Tseq{\Gamma}{\Thsubst CM}{\sigma}$.
\end{lemma}
The proof is a trivial induction on $C$.

\paragraph{Observational equivalence.}
Let $M,M'\in\Open\Gamma\sigma$ (that is, both terms have type $\sigma$ in the
typing context $\Gamma$). We say that $M$ and $M'$ are observationally
equivalent (notation $M\Rel\Obseq M'$) if, for all observation contexts $\Thole
C\Gamma\sigma$ such that $\Tseqh{}{C}{\Gamma}{\sigma}{\Tnat}$, one has
\begin{align*}
  \Redmat\Tnat^\infty_{\Thsubst CM,\Num 0}
  =\Redmat\Tnat^\infty_{\Thsubst C{M'},\Num 0}
\end{align*}


\begin{remark}
  The choice of testing the probability of reducing to $\Num 0$ in the
  definition above of observational equivalence is arbitrary. For instance, we
  would obtain the same notion of equivalence by stipulating that two terms $M$
  and $M'$ typable of type $\sigma$ in typing context $\Gamma$
  are observationally equivalent if, for all observation context
  $\Thole C\Gamma\sigma$,
  one has
  \begin{align*}
    \sum_{n\in\Nat}\Redmat\Tnat^\infty_{\Thsubst CM,\Num n}
    =\sum_{n\in\Nat}\Redmat\Tnat^\infty_{\Thsubst C{M'},\Num n}
  \end{align*}
  that is, the two closed terms $\Thsubst CM$ and $\Thsubst C{M'}$ have the same
  probability of convergence to some value. This is due to the universal
  quantification on $C$.
\end{remark}
\subsection{Basic examples}
\label{sec:basic-examples}

We give a series of terms written in $\PPCF$ which implement natural simple
algorithms to illustrate the expressive power of the language. We explain
intuitively the behavior of these programs, and one can also have a look at
\Parag{sec:interp-examples} where the denotational interpretations of
these terms in PCS are given, presented as functions.

Given a type $\sigma$, we set $\Omega_\sigma=\Fix{\Abst x\sigma x}$ so that
$\Tseq{}{\Omega_\sigma}\sigma$, which is the ever-looping term of type $\sigma$.

\paragraph{Arithmetics.}
The predecessor function, which is usually a basic construction of $\PCF$, is
now definable as:
\begin{equation*}
  \Pred=\Abst x\Tnat{\Ifv x{\Num 0}zz}
\end{equation*}
it is clear then that $\App\Pred{\Num 0}\Rel{\Transcl\Redoned}\Num 0$ and that
$\App\Pred{\Num{n+1}}\Rel{\Transcl\Redoned}\Num n$. 

The addition function can be defined as:
\begin{equation*}
  \Add=\Abst x\Tnat{\Fix{\Abst a{\Timpl\Tnat\Tnat}{\Abst
        y\Tnat{\Ifv{y}{x}{z}{\Succ{\App az}}}}}}
\end{equation*}
and it is easily checked that
$\Tseq{}{\Add}{\Timpl\Tnat{\Timpl\Tnat\Tnat}}$. 
Given $k\in\Nat$ we set 
\[
\Shift_k=\App\Add{\Num k}
\]
so that $\Tseq{}{\Shift_k}{\Timpl\Tnat\Tnat}$.

The exponential function can be defined as:
\begin{equation*}
  \Exp=\Fix{\Abst e{\Timpl\Tnat\Tnat}{\Abst x\Tnat{\Ifv
        x{\One}{z}{\Appp\Add{\App ez}{\,\App ez}}}}}
\end{equation*}
and satisfies $\App\Exp{\Num{n}}\Rel{\Transcl\Redoned}\Num{2^n}$.

In the same line, one defines a comparison function $\Cmp$
\begin{equation*}
  \Cmp=\Fix{\Abst c{\Timpl{\Tnat}{\Timpl\Tnat\Tnat}}{\Abst x{\Tnat}{\Abst
        y{\Tnat}{\Ifv{x}{\Num 0}{z}{\Ifv{y}{\Num 1}{z'}{\Appp cz{z'}}}}}}}
\end{equation*}
such that $\App{\Cmp}{\Num n\,\Num m}$ reduces to $\Num 0$ if $n\leq m$ and to
$\Num 1$ otherwise.

\paragraph{More tests.}
By induction on $k$, we define a family of terms $\Probe_k$ such that
$\Tseq{}{\Probe_k}{\Timpl\Tnat\Tnat}$:
\begin{align*}
  \Probe_0 &= \Abst x\Tnat{\Ifv{x}{\Num 0}{z}{\Omega_\Tnat}}\\
  \Probe_{k+1} &= \Abst x\Tnat{\Ifv{x}{\Omega_\Tnat}{z}{\App{\Probe_k}z}}
\end{align*}
For $M$ such that $\Tseq{}{M}{\Tnat}$, the term $\App{\Probe_k}M$ reduces to
$\Num 0$ with a probability which is equal to the probability of $M$ to reduce
to $\Num k$.

Similarly, we also define $\Pprod_k$ such that
$\Tseq{}{\Pprod_k}{\Timpl{\Tnat^k}{\Tnat}}$:
\begin{align*}
  \Pprod_0 &= \Num 0\\
  \Pprod_{k+1}
  &= \Abst x\Tnat{\Ifv x{\Pprod_k}{z}{\Omega_{\Timpl{\Tnat^k}{\Tnat}}}}\,.
\end{align*}
Given closed terms $\List M1k$ such that $\Tseq{}{M_i}\Tnat$, the term
$\App{\Pprod_k}{M_1\cdots M_k}$ reduces to $0$ with probability
$\prod_{i=1}^kp_i$ where $p_i$ is the probability of $M_i$ to reduce to $\Num
0$. 

Given a type $\sigma$ and $k\in\Nat$, we also define a term $\Pchoose_k$ such
that $\Tseq{}{\Pchoose_k}{\Timpl\Tnat{\Timpl{\sigma^k}{\sigma}}}$
\begin{align*}
  \Pchoose_0 &= \Abst\xi\Tnat{\Omega_\sigma}\\
  \Pchoose_{k+1} &=
  \Abstpref\xi\Tnat
  \Abstpref{x_1}\sigma\cdots\Abstpref{x_{k+1}}\sigma
  \Ifv\xi{x_1}{\zeta}{\Apppref{\Pchoose_k}{\zeta}
    \Argsep x_2\Argsep \cdots\Argsep  x_{k+1}}\,.
\end{align*}
Given a closed term $M$ such that $\Tseq{}{M}{\Tnat}$ and terms $\List N1k$
such that $\Tseq{\Gamma}{N_i}{\sigma}$ for each $i$, the term
$\App{\Pchoose_i}{M\Argsep N_1\cdots N_k}$ reduces to $N_i$ with the
probability that $M$ reduces to $\Num i$.

\paragraph{The let construction.}
This version of PCF, which is globally CBN, offers however the
possibility of handling integers in a CBV way. For instance, we can
set 
\begin{equation*}
  \Let xMN=\Ifv M{\Subst N{\Num 0}x}z{\Subst N{\Succ z}x}
\end{equation*}
and this construction can be typed as:
\begin{center}
  \AxiomC{$\Tseq\Gamma M\Tnat$}
  \AxiomC{$\Tseq{\Gamma,x:\Tnat}N\sigma$}
  \BinaryInfC{$\Tseq\Gamma{\Let xMN}\sigma$}
  \DisplayProof
\end{center}

One can also check that the following reduction inference holds
\begin{center}
  \AxiomC{$M\Rel{\Redone p}M'$}
  \UnaryInfC{$\Let xMN\Rel{\Redone p}\Let x{M'}N$}
  \DisplayProof
\end{center}
whereas \emph{it is no true} that
\begin{center}
  \AxiomC{$M\Rel{\Redone p}M'$}
  \UnaryInfC{$\Subst NMx\Rel{\Redone p}\Subst N{M'}x$}
  \DisplayProof  
\end{center}
(consider cases where $x$ does not occur in $N$, or occurs twice\dots). We have
of course
\begin{center}
  \AxiomC{}
  \UnaryInfC{$\Let x{\Num n}N\Rel\Redoned\Subst N{\theta(n)}x$}
  \DisplayProof
\end{center}
where $\theta(0)=\Num 0$ and $\theta({n+1})=\Succ{\Num n}$ (which
reduces to $\Num{n+1}$ in one deterministic step) by definition of this
construction.

\paragraph{Random generators.}
Using these constructions, we can define a closed term $\Unift$ of type
$\Timpl\Tnat\Tnat$ which, given an integer $n$, yields a uniform probability
distribution on the integers $0,\dots,2^n-1$:
\begin{equation*}
  \Unift=\Fix{\Abst u{\Timpl\Tnat\Tnat}{\Abst x\Tnat{\Ifv x{\Num
          0}{z}{\Ifv{\Dice{1/2}}{\App uz}{z'}{\Appp\Add{\App\Exp z}{\App
              uz}}}}}}\,.
\end{equation*}
Observe that, when evaluating $\App\Unift M$ (where $\Tseq{}{M}{\Tnat}$), the
term $M$ is evaluated only once thanks to the CBV feature of the conditional
construct. Indeed, we do not want the upper bound of the interval on which we
produce a probability distribution to change during the computation (the result
would be unpredictable!).

Using this construction, one can define a function $\Unif$ which, given an
integer $n$, yields a uniform probability distribution on the integers
$0,\dots,n $:
\begin{equation*}
  \Unif=\Abst x\Tnat{\Let yx{\Fix{\Abst u\Tnat{
          \Let z{\App\Unift y}{\Ifv{\Appp\Cmp zy}{z}{w}{\App uy}}}}}}
\end{equation*}
One checks easily that $\Tseq{}{\Unif}{\Timpl\Tnat\Tnat}$.  Given $n\in\Nat$,
this function applies iteratively $\Unift$ until the result is $\leq n$. It is
not hard to check that the resulting distribution is uniform (with probability
$\frac 1{n+1}$ for each possible result).

Last, let $n\in\Nat$ and let $\vec p=(\List p0n)$ be such that
$p_i\in[0,1]\cap\Rational$ and $p_0+\cdots+p_n\leq 1$. Then one defines a
closed term $\Ran{\vec p}$ which reduces to $\Num i$ with probability $p_i$ for
each $i\in\{0,\dots,n\}$. The definition is by induction on $n$.
\begin{equation*}
  \Ran{\List p0n}=
  \begin{cases}
    \Num 0 & \hspace{-10em}\text{if }p_0=1\text{ whatever be the value of }n\\
    \Ifv{\Dice{p_0}}{\Num 0}{z}{\Loopt\Tnat}& \hspace{-10em}\text{if }n=0\\
    \Ifv{\Dice{p_0}}
    {\Num 0}
    {z}{\Succ{\Ran{\frac{p_1}{1-p_0},\dots,\frac{p_n}{1-p_0}}}}
    &\text{otherwise}
  \end{cases}
\end{equation*}
Observe indeed that in the first case we must have $p_1=\dots=p_n=0$. 
%

\paragraph{A simple Las Vegas program.} 
Given a function $f:\Nat\to\Nat$ and $n\in\Nat$, find a $k\in\{0,\dots,n\}$
such that $f(k)=0$. This can be done by iterating random choices of $k$ until
we get a value such that $f(k)=0$: this is probably the simplest example of a
Las Vegas algorithm. The following function does the job:
\begin{align*}
  M=\Abst f{\Timpl\Tnat\Tnat}{\Abst x\Tnat{\Fix{\Abst r\Tnat{\Let{y}{\App\Unif
          x}{\Ifv{\App fy}{y}z{r}}}}}}
\end{align*}
with $\Tseq{}{M}{\Timpl{(\Timpl\Tnat\Tnat)}{\Timpl\Tnat\Tnat}}$. Our CBV
integers are crucial here since without our version of the conditional, it
would not be possible to get a random integer and use this value $y$ both as an
argument for $f$ and as a result if the expected condition holds.

%
%

\medbreak

We develop now a denotational semantics for this language.

\section{Probabilistic coherence spaces}\label{sec:PCS}
We present shortly a model of probabilistic PCF which is actually a model of
classical Linear Logic. For a longer and more detailed account, we refer
to~\cite{DanosEhrhard08}.

Let $I$ be a countable set. Given $u,u'\in\Realpto I$, we set 
\begin{align*}
  \Eval u{u'}=\sum_{i\in I}u_iu'_i\in\Realp\cup\{\infty\}\,.
\end{align*}
Let $\cX\subseteq\Realpto I$, we set
\begin{equation*}
  \Orth\cX=\{u'\in\Realpto I\St\forall u\in\cX\ \Eval u{u'}\leq 1\}\,.
\end{equation*}
We have as usual
\begin{itemize}
\item $\cX\subseteq\cY\Implies\Orth\cY\subseteq\Orth\cX$
\item $\cX\subseteq\Biorth\cX$
\end{itemize}
and it follows that $\Triorth\cX=\Orth\cX$.

\subsection{Definition and basic properties of probabilistic coherence spaces}
A \emph{probabilistic coherence space} (PCS) is a pair $X=(\Web X,\Pcoh X)$
where $\Web X$ is a countable set and $\Pcoh X\subseteq\Realpto{\Web X}$
satisfies
\begin{itemize}
\item $\Biorth{\Pcoh X}=\Pcoh X$ (equivalently, $\Biorth{\Pcoh X}\subseteq\Pcoh
  X$),
\item for each $a\in\Web X$ there exists $u\in\Pcoh X$ such that $u_a>0$,
\item for each $a\in\Web X$ there exists $A>0$ such that $\forall u\in\Pcoh X\
  u_a\leq A$.
\end{itemize}
If only the first of these conditions holds, we say that $X$ is a
\emph{pre-probabilistic coherence space} (pre-PCS).

The purpose of the second and third conditions is to prevent infinite
coefficients to appear in the semantics. This property in turn will be
essential for guaranteeing the morphisms interpreting proofs to be analytic
functions, which will be the key property to prove full abstraction. So these
conditions, though cosmetics at first sight, are important for our ultimate
goal.

\begin{lemma}\label{lemma:pcoh-charact}
  Let $X$ be a pre-PCS. The following conditions are
  equivalent:
  \begin{itemize}
  \item $X$ is a PCS,
  \item $\forall a\in\Web X\,\exists u\in\Pcoh X\,\exists
    u'\in\Orth{\Pcoh X}\ u_a>0\text{ and }u'_a>0$,
  \item $\forall a\in\Web X\,\exists A>0\,\forall u\in\Pcoh X\,\forall
    u'\in\Orth{\Pcoh X}\ u_a\leq A\text{ and }u'_a\leq A$.
  \end{itemize}
\end{lemma}
The proof is straightforward.

We equip $\Pcoh X$ with the most obvious partial order relation: $u\leq v$ if
$\forall a\in\Web X\ u_a\leq v_a$ (using the usual order relation on $\Real$).

Given $u\in\Realpto{\Web X}$ and $I\subseteq\Web X$ we use $\Restr uI$ for the
element $v$ of $\Realpto{\Web X}$ such that $v_a=u_a$ if $a\in I$ and
$v_a=0$ otherwise. Of course $u\in\Pcoh X\Implies\Restr uI\in\Pcoh X$.

\begin{theorem}\label{th:Pcoh-prop}
  $\Pcoh X$ is an $\omega$-continuous domain. Given $u,v\in\Pcoh X$ and
  $\alpha,\beta\in\Realp$ such that $\alpha+\beta\leq 1$, one has $\alpha
  u+\beta v\in\Pcoh X$.
\end{theorem}
\Beginproof
Let us first prove that $\Pcoh X$ is complete. Let $D$ be a directed subset of
$\Pcoh X$. For any $a\in\Web X$, the set $\{u_a\St u\in D\}$ is bounded; let
$v_a\in\Realp$ be the lub of that set. In that way we define $v=(v_a)_{a\in\Web
X}\in\Realpto{\Web X}$. 

We prove that $v\in\Pcoh X$. Let $u'\in\Orth{\Pcoh X}$, we must prove that
$\Eval v{u'}\leq 1$. We know that $\{\Eval u{u'}\St u\in D\}\subseteq[0,1]$ and
therefore this set has a lub $A\in[0,1]$. Let $\epsilon>0$, we can find $u\in
D$ such that $\Eval u{u'}\geq A-\epsilon$ and since this holds for all
$\epsilon$, we have $\Eval v{u'}\geq A$. Let again $\epsilon>0$. We can find a
finite set $I\subseteq\Web X$ such that $\Eval{\Restr vI}{u'}\geq\Eval
v{u'}-\frac\epsilon 2$. Since $I$ is finite we have $\Eval{\Restr
  vI}{u'}=\sup_{u\in D}\Eval{\Restr uI}{u'}$ (it is here that we use our
hypothesis that $D$ is directed) and hence we can find $u\in D$
such that $\Eval{\Restr uI}{u'}\geq\Eval{\Restr vI}{u'}-\frac{\epsilon}{2}$ and
hence 
$\Eval u{u'}\geq\Eval{\Restr uI}{u'}
\geq\Eval{\Restr vI}{u'}-\frac{\epsilon}{2}\geq\Eval v{u'}-\epsilon$. It
follows that $A=\sup_{u\in D}\Eval u{u'}\geq\Eval v{u'}$. So $\Eval
v{u'}\in[0,1]$ and hence $v\in\Pcoh X$.

It is clear that $v$ is the lub of $D$ in $\Pcoh X$ since
the order relation is defined pointwise. Therefore $\Pcoh X$ is a cpo, which
has $0$ as least element. Let $R$ be the set of all the elements of $\Pcoh X$
which have a finite domain and take only rational values. Then it is clear that
for each $u\in\Pcoh X$, the elements $w$ of $R$ which are way below $u$ (in the
present setting, this simply means that $w_a>0\Implies w_a<u_a$) form a
directed subset of $\Pcoh X$ whose lub is $u$. Therefore $\Pcoh X$ is an
$\omega$-continuous domain.

The last statement results from the linearity of the operation $u\mapsto\Eval
u{u'}$. 
\Endproof

As a consequence, given a family $(u(i))_{i\in\Nat}$ of elements of $\Pcoh X$
and a family $(\alpha_i)_{i\in\Nat}$ of elements of $\Realp$ such that
$\sum_{i\in\Nat}\alpha_i\leq 1$, one has $\sum_{i\in\Nat}\alpha_iu(i)\in\Pcoh
X$. 

\subsection{Morphisms of PCSs}

Let $X$ and $Y$ be PCSs. Let $t\in(\Realp)^{\Web X\times\Web Y}$ (to be
understood as a matrix). Given $u\in\Pcoh X$, we define $\Matapp
tu\in\Realpc^{\Web Y}$ by $(\Matapp tu)_b=\sum_{a\in\Web X}t_{a,b}u_a$
(application of the matrix $t$ to the vector $u$)\footnote{This is an unordered
  sum, which is infinite in general. It makes sense because all its terms are
  $\geq 0$.}.  We say that $t$ is a \emph{(linear) morphism} from $X$ to $Y$ if
$\forall u\in\Pcoh X\ \Matapp tu\in\Pcoh Y$, that is
\begin{align*}
  \forall u\in\Pcoh X\,\forall{v'}\in\Orth{\Pcoh Y}\quad\sum_{(a,b)\in\Web
    X\times\Web Y}t_{a,b}u_av'_b\leq 1\,.
\end{align*}
The diagonal matrix $\Id\in(\Realp)^{\Web X\times\Web X}$ given by
$\Id_{a,b}=1$ if $a=b$ and $\Id_{a,b}=0$ otherwise is a morphism. In that way
we have defined a category $\PCOH$ whose objects are the PCSs and whose
morphisms have just been defined. Composition of morphisms is defined as matrix
multiplication: let $s\in\PCOH(X,Y)$ and $t\in\PCOH(Y,Z)$, we define $\Matapp
ts\in(\Realp)^{\Web X\times\Web Z}$ by
\begin{align*}
  (\Matapp ts)_{a,c}=\sum_{b\in\Web Y}s_{a,b}t_{b,c}
\end{align*}
and a simple computation shows that $\Matapp ts\in\PCOH(X,Z)$. More precisely,
we use the fact that, given $u\in\Pcoh X$, one has $\Matapp{(\Matapp
  ts)}{u}=\Matapp t{(\Matapp su)}$. Associativity of composition holds because
matrix multiplication is associative. $\Id_X$ is the identity morphism at $X$.

\subsection{The norm}
Given $u\in\Pcoh X$, we define $\Norm u_X=\sup\{\Eval u{u'}\St u'\in\Pcoh{\Orth
X}\}$. By definition, we have $\Norm u_X\in[0,1]$.

\subsection{Multiplicative constructs}\label{sec:multiplicatives}
We start the description of the category $\PCOH$ as a model of Linear
Logic. For this purpose, we use Bierman's notion of Linear
Category~\cite{Bierman95}, as presented in~\cite{Mellies09} which is our main
reference for this topic. 

One sets $\Orth X=(\Web X,\Orth{\Pcoh X})$. It results straightforwardly from
the definition of PCSs that $\Orth X$ is a PCS. Given $t\in\PCOH(X,Y)$, one has
$\Orth t\in\PCOH(\Orth Y,\Orth X)$ if $\Orth t$ is the transpose of $t$, that
is $(\Orth t)_{b,a}=t_{a,b}$.

One defines $\Tens{X}{Y}$ by
$\Web{\Tens{X}{Y}}=\Web{X}\times\Web{Y}$ and
\begin{equation*}
  \Pcohp{\Tens XY}=\Biorth{\{\Tens uv\St u\in\Pcoh X\text{ and }v\in\Pcoh Y\}}
\end{equation*}
where $\Tensp uv_{(a,b)}=u_av_b$.  Then $\Tens XY$ is a pre-PCS.

We have 
\[
\Pcoh{\Orth{(\Tens{X}{\Orth Y})}}=\Orth{\{\Tens{u}{v'}\St u\in\Pcoh X\text{ and
  }v'\in\Pcoh{\Orth Y}\}}=\PCOH(X,Y)\,.
\]
It follows that $\Limpl XY=\Orth{(\Tens{X}{\Orth Y})}$ is a pre-PCS. Let
$(a,b)\in\Web X\times\Web Y$. Since $X$ and $\Orth Y$ are PCSs, there is $A>0$
such that $u_av'_b<A$ for all $u\in\Pcoh X$ and $v'\in\Pcoh{\Orth Y}$. Let
$t\in(\Realp)^{\Web{\Limpl XY}}$ be such that $t_{(a',b')}=0$ for
$(a',b')\not=(a,b)$ and $t_{(a,b)}=1/A$, we have $t\in\Pcoh{(\Limpl XY)}$. This
shows that $\exists t\in\Pcoh{(\Limpl XY)}$ such that $t_{(a,b)}>0$. Similarly
we can find $u\in\Pcoh X$ and $v'\in\Pcoh{\Orth Y}$ such that
$\epsilon=u_av'_b>0$. It follows that $\forall t\in\Pcoh{(\Limpl XY)}$ one has
$t_{(a,b)}\leq 1/\epsilon$. We conclude that $\Limpl XY$ is a PCS, and
therefore $\Tens XY$ is also a PCS.

\begin{lemma}\label{lemma:PCS-moprh-charact}
  Let $X$ and $Y$ be PCSs. One has $\Pcoh{(\Limpl XY)}=\PCOH(X,Y)$. That is,
  given $t\in\Realpto{\Web X\times\Web Y}$, one has $t\in\Pcoh{(\Limpl XY)}$
  iff for all $u\in\Pcoh X$, one has $t\Compl u\in\Pcoh Y$.
\end{lemma}
This results immediately from the definition above of $\Limpl XY$.

\begin{lemma}\label{lemma:tens-morph-charact}
  Let $X_1$, $X_2$ and $Y$ be PCSs. Let
  $t\in(\Realp)^{\Web{\Limpl{\Tens{X_1}{X_2}}{Y}}}$. One has
  $t\in\PCOH(\Tens{X_1}{X_2},Y)$ iff for all $u_1\in\Pcoh{X_1}$ and
  $u_2\in\Pcoh{X_2}$ one has $\Matapp t{(\Tens{u_1}{u_2})}\in\Pcoh Y$.
\end{lemma}
\Beginproof
The condition stated by the lemma is clearly necessary. Let us prove that it is
sufficient: under this condition, it suffices to prove that
\begin{align*}
  \Orth t\in\PCOH(\Orth Y,\Orth{(\Tens{X_1}{X_2})})\,.
\end{align*}
Let $v'\in\Pcoh{\Orth Y}$, it suffices to prove that $\Matapp{\Orth
  t}{v'}\in\Pcoh{\Orth{(\Tens{X_1}{X_2})}}$. So let $u_1\in\Pcoh{X_1}$ and
$u_2\in\Pcoh{X_2}$, it suffices to prove that $\Eval{\Matapp{\Orth
    t}{v'}}{\Tens{u_1}{u_2}}\leq 1$, that is $\Eval{\Matapp
  t{(\Tens{u_1}{u_2})}}{v'}\leq 1$, which follows from our assumption.
\Endproof

Let $s_i\in\PCOH(X_i,Y_i)$ for $i=1,2$. Then one defines
\begin{align*}
\Tens{s_1}{s_2}\in(\Realp)^{\Web{\Limpl{\Tens{X_1}{X_2}}{\Tens{Y_1}{Y_2}}}}  
\end{align*}
by
$(\Tens{s_1}{s_2})_{((a_1,a_2),(b_1,b_2))}=(s_1)_{(a_1,b_1)}(s_2)_{(a_2,b_2)}$
and one must check that
$\Tens{s_1}{s_2}\in\PCOH({\Tens{X_1}{X_2},\Tens{Y_1}{Y_2}})$. This follows
directly from Lemma~\ref{lemma:tens-morph-charact}. Let
$\One=(\{*\},[0,1])$. There are obvious choices of natural isomorphisms
\begin{align*}
  \Leftu_X&\in\PCOH(\Tens\One X,X)\\
  \Rightu_X&\in\PCOH(\Tens X\One,X)\\
  \Assoc_{X_1,X_2,X_3}&\in\PCOH(\Tens{\Tensp{X_1}{X_2}}{X_3},
      \Tens{X_1}{\Tensp{X_2}{X_3}})\\
  \Sym_{X_1,X_2}&\in\PCOH(\Tens{X_1}{X_2},\Tens{X_2}{X_1})  
\end{align*}
which satisfy the standard coherence properties. This shows that the structure
$(\PCOH,\One,\Leftu,\Rightu,\Assoc,\Sym)$ is a symmetric monoidal category.

\paragraph{Internal linear hom.}
Given PCSs $X$ and $Y$, let us define
$\Evlin\in(\Realp)^{\Web{\Limpl{\Tens{(\Limpl XY)}{X}}{Y}}}$ by
\begin{align*}
\Evlin_{(((a',b'),a),b)}=
\begin{cases}
  1 & \text{if }(a,b)=(a',b')\\
  0 & \text{otherwise.}
\end{cases}
\end{align*}
Then it is easy to see that $(\Limpl XY,\Evlin)$ is an internal linear hom
object in $\PCOH$, showing that this SMCC is closed. If $t\in\PCOH(\Tens
ZX,Y)$, the corresponding linearly curryfied morphism
$\Curlin(t)\in\PCOH(Z,\Limpl XY)$ is given by
$\Curlin(t)_{(c,(a,b))}=t_{((c,a),b)}$. 

\paragraph{*-autonomy.}
Take $\Bot=\One$, then one checks readily that the structure
$(\PCOH,\One,\Leftu,\Rightu,\Assoc,\Sym,\Bot)$ is a *-autonomous category. The
duality functor $X\mapsto(\Limpl X\Bot)$ can be identified with the strictly
involutive contravariant functor $X\mapsto\Orth X$.

\subsection{Additives}
Let $(X_i)_{i\in I}$ be a countable family of PCSs. We define a PCS
$\Bwith_{i\in I}X_i$ by $\Web{\Bwith_{i\in I}X_i}=\bigcup_{i\in
  I}\{i\}\times\Web{X_i}$ and $u\in\Pcohp{\Bwith_{i\in I}X_i}$ if, for all
$i\in I$, the family $u(i)\in(\Realp)^{\Web{X_i}}$ defined by
$u(i)_a=u_{(i,a)}$ belongs to $\Pcoh{X_i}$.

\begin{lemma}
  Let $u'\in(\Realp)^{\Web{\Bwith_{i\in I}X_i}}$. One has
  $u'\in\Pcoh{\Orthp{\Bwith_{i\in I}X_i}}$ iff
  \begin{itemize}
  \item $\forall i\in I\ u'(i)\in\Pcoh{\Orth{X_i}}$
  \item and $\sum_{i\in I}\Norm{u'(i)}_{\Orth{X_i}}\leq 1$.
  \end{itemize}
\end{lemma}
The proof is quite easy. It follows that $\Bwith_{i\in I}X_i$ is a
PCS. Moreover we can define $\Proj i\in\PCOH(\Bwith_{j\in I}X_j,X_i)$ by
\begin{align*}
(\Proj i)_{(j,a),a'}=
\begin{cases}
  1 & \text{if }j=i\text{ and }a=a'\\
  0 & \text{otherwise.}
\end{cases}
\end{align*}
Then $(\Bwith_{i\in I}X_i,(\Proj i)_{i\in I})$ is the cartesian product of the
family $(X_i)_{i\in I}$ in the category $\PCOH$. The coproduct $(\Bplus_{i\in
  I}X_i,(\Inj i)_{i\in I})$ is the dual operation, so that
\begin{align*}
  \Web{\Bplus_{i\in I}X_i}=\Union_{i\in I}\{i\}\times\Web{X_i}
\end{align*}
and $u\in\Pcoh{(\Bplus_{i\in I}X_i)}$ if $\forall i\in I\ u(i)\in\Pcoh{X_i}$
and $\sum_{i\in I}\Norm{u(i)}_{X_i}\leq 1$. The injections
$\Inj j\in\PCOH(X_j,\Bplus_{i\in I}X_i)$ are given by
\begin{align*}
(\Inj i)_{a',(j,a)}=
\begin{cases}
  1 & \text{if }j=i\text{ and }a=a'\\
  0 & \text{otherwise.}
\end{cases}
\end{align*}

We define in particular $\Pnat=\Bplus_{i\in\Nat}\One$, that is
$\Web\Pnat=\Nat$ and $u\in(\Realp)^\Nat$ belongs to $\Pcoh\Pnat$ if
$\sum_{n\in\Nat}u_n\leq 1$.  

\subsection{Exponentials}
Given a set $I$, a \emph{finite multiset} of elements of $I$ is a function
$\mu:I\to\Nat$ whose \emph{support} $\Supp\mu=\{a\in I\St\mu(a)\not=0\}$ is
finite. We use $\Mfin I$ for the set of all finite multisets of elements of
$I$. Given a finite family $\List a1n$ of elements of $I$, we use $\Mset{\List
  a1n}$ for the multiset $\mu$ such that $\mu(a)=\Card{\{i\St a_i=a\}}$. We use
additive notations for multiset unions: $\sum_{i=1}^k\mu_i$ is the multiset
$\mu$ such that $\mu(a)=\sum_{i=1}^k\mu_i(a)$. The empty multiset is denoted as
$0$ or $\Msetempty$. If $k\in\Nat$, the multiset $k\mu$ maps $a$ to $k\mu(a)$.

Let $X$ be a PCS. Given $u\in\Pcoh X$ and $\mu\in\Mfin{\Web X}$, we define
$u^\mu=\prod_{a\in\Web X}u_a^{\mu(a)}\in\Realp$. Then we set $\Prom
u=(u^\mu)_{\mu\in\Mfin{\Web X}}$ and finally
\begin{align*}
  \Excl X=(\Mfin{\Web X},\Biorth{\{\Prom u\St u\in\Pcoh X\}})
\end{align*}
which is a pre-PCS. 

We check quickly that $\Excl X$ so defined is a PCS. Let $\mu=\Mset{\List
  a1n}\in\Mfin{\Web X}$.  Because $X$ is a PCS, and by
Theorem~\ref{th:Pcoh-prop}, for each $i=1,\dots,n$ there is $u(i)\in\Pcoh X$
such that $u(i)_{a_i}>0$. Let $(\alpha_i)_{i=1}^n$ be a family of strictly
positive real numbers such that $\sum_{i=1}^n\alpha_i\leq 1$. Then
$u=\sum_{i=1}^n\alpha_iu(i)\in\Pcoh X$ satisfies $u_{a_i}>0$ for each
$i=1,\dots,n$. Therefore $\Prom u_\mu=u^\mu>0$. This shows that there is
$U\in\Pcoh{(\Excl X)}$ such that $U_\mu>0$.

Let now $A\in\Realp$ be
such that $\forall u\in\Pcoh X\,\forall i\in\{1,\dots,n\}\ u_{a_i}\leq A$. For
all $u\in\Pcoh X$ we have $u^\mu\leq A^n$. We have
\begin{align*}
  \Orth{(\Pcoh{(\Excl X)})}=\Triorth{\{\Prom u\St u\in\Pcoh X\}}
  =\Orth{\{\Prom u\St u\in\Pcoh X\}}\,.
\end{align*}
Let $t\in\Realpto{\Web{\Excl X}}$ be defined by $t_\nu=0$ if $\nu\not=\mu$ and
$t_\mu=A^{-n}>0$; we have $t\in\Orth{(\Pcoh{(\Excl X)})}$. We have exhibited an
element $t$ of $\Orth{(\Pcoh{(\Excl X)})}$ such that $t_\mu>0$. By
Lemma~\ref{lemma:pcoh-charact} it follows that $\Excl X$ is a PCS.

\paragraph{Kleisli morphisms as functions.}
Let $s\in\Realpto{\Web{\Limpl{\Excl X}{Y}}}$. We define a function $\Fun
s:\Pcoh X\to\Realpcto{\Web Y}$ as follows.  Given $u\in\Pcoh X$, we set
\begin{align*}
  \Fun s(u)=s\Compl {\Prom u}=\left(\sum_{\mu\in\Web{\Excl
        X}}s_{\mu,b}u^\mu\right)_{b\in\Web Y}\,.
\end{align*}

\begin{proposition}\label{prop:kleisli-morph-charact}
  One has $s\in\Pcoh{(\Limpl{\Excl X}{Y})}$ iff, for all $u\in\Pcoh X$, one has
  $\Fun s(u)\in\Pcoh Y$.  
\end{proposition}
\Beginproof
By Lemma~\ref{lemma:PCS-moprh-charact}, the condition is necessary since
$u\in\Pcoh X\Implies\Prom u\in\Pcoh{(\Excl X)}$, let us prove that it is
sufficient. Given $v'\in\Orth{\Pcoh Y}$, it suffices to prove that $\Orth
s\Compl v'\in\Orth{\Pcoh{(\Excl X)}}$, that is $\Eval{\Orth s\Compl v'}{\Prom
  u}\leq 1$ for all $u\in\Pcoh X$. This results from the assumption because
$\Eval{\Orth s\Compl v'}{\Prom u}=\Eval{\Fun s(u)}{v'}$.
\Endproof

\begin{theorem}\label{th:pcoh-functional}
  Let $s\in\PCOH(\Excl X,Y)$.  The function $\Fun s$ is
  Scott-continuous. Moreover, given $s,s'\in\PCOH(\Excl X,Y)$, one has $s=s'$
  (as matrices) iff $\Fun s=\Fun{s'}$ (as functions $\Pcoh X\to\Pcoh Y$).
\end{theorem}
\Beginproof
%
%
Let us first prove that $\Fun s$ is Scott continuous. It is clear that this
function is monotone. Let $D$ be a directed subset of $\Pcoh X$ and let $w$ be
its lub, we must prove that $\Fun s(w)=\sup_{u\in D}{\Fun s(u)}$. Let $b\in\Web
Y$. Since multiplication is a Scott-continuous function from $\Rseg 01^2$ to
$\Rseg 01$, we have $\Fun s(w)_b=\sum_{\mu\in\Web{\Excl X}}\sup_{u\in
  D}s_{\mu,b}u^\mu$. The announced property follows by the monotone convergence
theorem.

Let now $s,s'\in\PCOH(\Excl X,Y)$ be such that $\Fun s(u)=\Fun{s'}(u)$ for all
$u\in\Pcoh X$. Let $\mu\in\Web{\Excl X}$ and $b\in\Web Y$, we prove that
$s_{\mu,b}=s'_{\mu,b}$. Let $I=\Supp\mu$. Given $u\in\Realpto I$, let
$\eta(u)\in\Realpto{\Web X}$ be defined by $\eta(u)_a=0$ for $a\notin u$ and
$\eta(u)_a=u_a$ for $a\in I$. Let $A>0$ be such that
$\eta([0,A]^I)\subseteq\Pcoh X$ (such an $A$ exists because $I$ is finite and
by our definition of PCS). Let $\Proj b:\Pcoh Y\to\Real$ be defined by $\Proj
b(v)=v_b$. Let $f=\Proj b\Comp\Fun s\Comp\eta:[0,A]^I\to\Real$ and $f'=\Proj
b\Comp\Fun{s'}\Comp\eta$. Then we have $f=f'$ by our assumption on $s$ and
$s'$. But $f$ and $f'$ are entire functions and we have $f(u)=\sum_{\nu\in\Mfin
  I}s_{\nu,b}\prod_{a\in I}u_a^{\nu(a)}$ and similarly for $f'$ and $s'$. Since
$[0,A]^I$ contains a non-empty open subset of $\Real^I$, it follows that
$s_{\nu,b}=s'_{\nu,b}$ for all $\nu\in\Mfin I$. In particular,
$s_{\mu,b}=s'_{\mu,b}$.
\Endproof

So we can consider the elements of $\Kl\PCOH(X,Y)$ (the morphisms of the
Kleisli category of the comonad $\Excl\_$ on the category $\PCOH$) as
particular Scott continuous functions $\Pcoh X\to\Pcoh Y$. Of course, not all
Scott continuous function are morphisms in $\Kl\PCOH$.

\begin{example}
  Take $X=Y=\One$. A morphism in $\Kl\PCOH(\One,\One)$ can be seen as a
  function $f:[0,1]\to[0,1]$ such that $f(u)=\sum_{n=0}^\infty s_nu^n$ where
  the $s_n$'s are $\geq 0$ and satisfy $\sum_{n=0}^\infty s_n\leq 1$. Of
  course, not all Scott continuous function $[0,1]\to[0,1]$ are of that
  particular shape! Take for instance the function $f:[0,1]\to[0,1]$ defined by
  $f(u)=0$ if $u\leq\frac 12$ and $f(u)=2u-1$ if $u>\frac 12$; this function
  $f$ is Scott continuous but has no derivative at $u=\frac 12$ and therefore
  cannot be expressed as a power series.
\end{example}

\begin{proposition}\label{prop:order-fun-kleiseli}
  Let $s,s'\in\Kl\PCOH(X,Y)$ be such that $s\leq s'$ (as elements of
  $\Pcoh{(\Limpl{\Excl X}Y)}$). Then $\forall u\in\Pcoh X\ \Fun
  s(u)\leq\Fun{s'}(u)$. Let $(s(i))_{i\in\Nat}$ be a monotone sequence of
  elements of $\Kl\PCOH(X,Y)$ and let $s=\sup_{i\in\Nat}s(i)$. Then $\forall
  u\in\Pcoh X\ \Fun s(u)=\sup_{i\in I}\Fun{s_i}(u)$.
\end{proposition}
\Beginproof
The first statement is obvious. The second one results from the monotone
convergence Theorem.
\Endproof


\begin{remark}
  We can have $s,s'\in\Kl\PCOH(X,Y)$ such that $\forall u\in\Pcoh X\ \Fun
  s(u)\leq\Fun{s'}(u)$ but without having that $s\leq s'$. Take for instance
  $X=Y=\One$.
  As in the example above we can see $\Fun s$ and $\Fun{s'}$ as functions
  $[0,1]\to[0,1]$ given by $\Fun s(u)=\sum_{n=0}^\infty s_nu^n$ and
  $\Fun{s'}(u)=\sum_{n=0}^\infty s'_nu^n$, and $s\leq s'$ means that $\forall
  n\in\Nat\ s_n\leq s'_n$. Then let $s$ be defined by $s_n=1$ if $n=2$ and
  $s_n=0$ otherwise, and $s'$ be defined by $s'_n=1$ if $n=1$ and $s'_n=0$
  otherwise. We have $\Fun s(u)=u^2\leq\Fun{s'}(u)=u$ for all $u\in[0,1]$
  whereas $s$ and $s'$ are not comparable in
  $\Pcoh{(\Limpl{\Excl\One}{\One})}$.
\end{remark}

Given a multiset $\mu\in\Mfin I$, we define its \emph{factorial}
$\Factor\mu=\prod_{i\in I}\Factor{\mu(i)}$ and its \emph{multinomial
  coefficient} $\Multinom{}\mu=\Factor{(\Card\mu)}/\Factor\mu\in\Natnz$ where
$\Card\mu=\sum_{i\in I}\mu(i)$ is the cardinality of $\mu$. Remember that,
given an $I$-indexed family $a=(a_i)_{i\in I}$ of elements of a commutative
semi-ring, one has the multinomial formula
\begin{align*}
  \Big(\sum_{i\in I}a_i\Big)^n=\sum_{\mu\in\Mfinc nI}\Multinom{}\mu a^\mu
\end{align*}
where $\Mfinc nI=\{\mu\in\Mfin I\St\Card\mu=n\}$.

Given $\mu\in\Web{\Excl X}$ and $\nu\in\Web{\Excl Y}$ we define
$\Mexpset\mu\nu$ as the set of all multisets $\rho$ in $\Mfin{\Web X\times\Web
  Y}$ such that
\begin{align*}
  \forall a\in\Web X\ \sum_{b\in\Web Y}\rho(a,b)=\mu(a)
  \quad\text{and}\quad
  \forall b\in\Web Y\ \sum_{a\in\Web X}\rho(a,b)=\nu(b)\,.
\end{align*}

Let $t\in\PCOH(X,Y)$, we define $\Excl t\in\Realpto{\Limpl{\Excl X}{\Excl Y}}$
by
\begin{align*}
  (\Excl t)_{\mu,\nu}
  =\sum_{\rho\in\Mexpset\mu\nu}\frac{\Factor\nu}{\Factor\rho}t^\rho\,.
\end{align*}
Observe that the coefficients in this sum are all non-negative integers.

\begin{lemma}\label{lemma:excl-morph-app}
  For all $u\in\Pcoh X$ one has $\Excl t\Compl\Prom u=\Prom{(t\Compl u)}$.
\end{lemma}
\Beginproof
Indeed, given $\nu\in\Web{\Excl Y}$, one has
\begin{align*}
  \Prom{(t\Compl u)}_\nu
  &= \prod_{b\in\Web Y}\Big(\sum_{a\in\Web X}t_{a,b}u_a\Big)^{\nu(b)}\\
  &= \prod_{b\in\Web Y}\Bigg(\sum_{
           \overset{\scriptstyle{\mu\in\Web{\Excl X}}}{\Card\mu=\nu(b)}}
         \Multinom{}\mu u^\mu\prod_{a\in\Web X}t_{a,b}^{\mu(a)}\Bigg)\\
  &= \sum_{
         \overset{\scriptstyle{\theta\in\Web{\Excl X}^{\Web Y}}}
                 {\forall b\ \Card{\theta(b)}=\nu(b)}}
         u^{\sum_{b\in\Web{Y}}\theta(b)}
         \Big(\prod_{b\in\Web Y}\Multinom{}{\theta(b)}\Big)
         \Bigg(\prod_{\Biind{a\in\Web X}
           {b\in\Web Y}}t_{a,b}^{\theta(b)(a)}\Bigg)\\
  &= \sum_{\mu\in\Web{\Excl X}} u^\mu
     \sum_{\rho\in\Mexpset\mu\nu}t^\rho\prod_{b\in Y}
       \frac{\Factor{\nu(b)}}{\prod_{a\in\Web X}\Factor{\rho(a,b)}}\\
  &= \sum_{\mu\in\Web{\Excl X}}(\Excl t)_{\mu,\nu}u^\mu
\end{align*}
since there is a bijective correspondence between the $\theta\in\Web{\Excl
  X}^{\Web Y}$ such that $\forall b\in\Web Y\ \Card{\theta(b)}=\nu(b)$ and the
$\rho\in\bigcup_{\mu\in\Web{\Excl X}}\Mexpset\mu\nu$ (observe that this union
of sets is actually a disjoint union): this bijection maps $\theta$ to the
multiset $\rho$ defined by $\rho(a,b)=\theta(b)(a)$.
\Endproof

\begin{proposition}
  For all $t\in\PCOH(X,Y)$ one has $\Excl t\in\PCOH(\Excl X,\Excl Y)$ and the
  operation $t\mapsto\Excl t$ is functorial.
\end{proposition}
\Beginproof
Immediate consequences of Lemma~\ref{lemma:excl-morph-app} and
Theorem~\ref{th:pcoh-functional}. 
\Endproof

\paragraph{Description of the exponential comonad.}
We equip now this functor with a structure of comonad: let $\Der
X\in\Realpto{\Web{\Limpl{\Excl X}X}}$ be given by $(\Der
X)_{\mu,a}=\Kronecker{\Mset a}{\mu}$ (the value of the Kronecker symbol
$\Kronecker ij$ is $1$ if $i=j$ and $0$ otherwise) and $\Digg
X\in\Realpto{\Web{\Limpl{\Excl X}{\Excl{\Excl X}}}}$ be given by $(\Digg
X)_{\mu,\Mset{\List\mu 1n}}=\Kronecker{\sum_{i=1}^n\mu_i}{\mu}$. Then we have
$\Der X\in\PCOH(\Excl X,X)$ and $\Digg X\in\PCOH(\Excl X,\Excl{\Excl X})$
simply because
\begin{align*}
  \Fun{\Der X}(u)=u\quad\text{and}\quad\Fun{\Digg X}(u)=\Prom{(\Prom u)}
\end{align*}
for all $u\in\Pcoh X$, as easily checked. Using these equations, one also
checks easily the naturality of these morphisms, and the fact that
$(\Excl\_,\Der{},\Digg{})$ is a comonad.

As to the monoidality of this comonad, we introduce
$\Expmonisoz\in\Realpto{\Web{\Limpl{\One}{\Excl\Top}}}$ by
$\Expmonisoz_{*,\Mset{}}=1$ and $\Expmonisob
XY\in\Realpto{\Web{\Limpl{\Tens{\Excl X}{\Excl Y}}{\Excl{(\With XY)}}}}$ by
$(\Expmonisob XY)_{\lambda,\rho,\mu}=\Kronecker{\mu}{\Injms 1\lambda+\Injms
  2\rho}$ 
where $\Injms i{\Mset{\List a1n}}=\Mset{(i,a_1),\dots,(i,a_n)}$. It is easily
checked that the required commutations hold (again, we refer
to~\cite{Mellies09}).

It follows that we can define a lax symmetric monoidal structure for the
functor $\Excl\_$ from the symmetric monoidal category $(\PCOH,\ITens)$ to
itself, that is, for each $n\in\Nat$, a natural morphism
\[
\Monoidal^{(n)}_{\List X1N}\in\PCOH(\Excl{X_1}\ITens\cdots\ITens\Excl{X_n},
\Excl{(X_1\ITens\cdots\ITens X_n)})
\]
satisfying some coherence conditions.

Given $f\in\PCOH(\Excl{X_1}\ITens\cdots\ITens\Excl{X_n},Y)$, we define the
\emph{promotion} morphism $\Prom
f\in\PCOH(\Excl{X_1}\ITens\cdots\ITens\Excl{X_n},\Excl Y)$ as the following
composition of morphisms in $\PCOH$
\begin{align}
  \begin{tikzpicture}[baseline=(current  bounding  box.center), ->, >=stealth]
    \node (1) {$\Excl{X_1}\ITens\cdots\ITens\Excl{X_n}$};
    \node (2) [ below of = 1, node distance = 1cm ]
      {$\Excl{\Excl{X_1}}\ITens\cdots\ITens\Excl{\Excl{X_n}}$};
    \node (3) [ right of = 2, node distance = 5cm ]
      {$\Excl{(\Excl{X_1}\ITens\cdots\ITens\Excl{X_n})}$};
    \node (4) [ above of = 3, node distance = 1cm ]
      {$\Excl Y$};
    \tikzstyle{every node}=[midway,auto,font=\scriptsize]
    \draw (1) -- node [swap] {$\Digg{X_1}\ITens\cdots\ITens\Digg{X_n}$} (2);
    \draw (2) -- node {$\Monoidal^{(n)}_{\Excl{X_1},\dots,\Excl{X_n}}$} (3);
    \draw (3) -- node [swap] {$\Excl f$} (4);
  \end{tikzpicture}
  \label{eq:promo-def}
\end{align}

\paragraph{Cartesian closeness of the Kleisli category.}
The Kleisli category $\Kl\PCOH$ of the comonad $\Excl\_$ has the same objects
as $\PCOH$, and $\Kl\PCOH(X,Y)=\PCOH(\Excl X,Y)$. The identity morphism at
object $X$ is $\Der X$ and given $f\in\Kl\PCOH(X,Y)$ and $g\in\Kl\PCOH(Y,Z)$
the composition of $f$ and $g$ in $\Kl\PCOH$, denoted as $g\Comp f$, is given by
\begin{align*}
  g\Comp f=g\Compl\Prom f\,.
\end{align*}

This category is cartesian closed: the terminal object is $\Top$, the cartesian
product of two objects $X$ and $Y$ is $\With XY$ (with projections defined in
the obvious way, using $\Der{\With XY}$ and the projections of the cartesian
product in $\PCOH$), their internal hom object is $\Impl XY=\Limpl{\Excl
  X}{Y}$. The corresponding evaluation morphism $\Ev\in\Kl\PCOH(\With{(\Impl
  XY)}{X},Y)$ is defined as the following composition of morphisms in $\PCOH$
\begin{center}
  \begin{tikzpicture}[->, >=stealth]
    \node (1) {$\Excl{(\With{(\Impl XY)}{X})}$};
    \node (2) [right of=1, node distance=3.6cm]
      {$\Tens{\Excl{(\Impl XY)}}{\Excl X}$};
    \node (3) [right of=2, node distance=3.6cm]
      {$\Tens{{(\Impl XY)}}{\Excl X}$};
    \node (4) [right of=3, node distance=2cm] {$Y$};
    \tikzstyle{every node}=[midway,auto,font=\scriptsize]
    \draw (1) -- node {$\Funinv{(\Expmonisobn)}$} (2);
    \draw (2) -- node {$\Tens{\Der{}}{\Excl X}$} (3);
    \draw (3) -- node {$\Evlin$} (4);
  \end{tikzpicture}  
\end{center}

The curryfied version of a morphism $t\in\Kl\PCOH(\With ZX,Y)$ is the morphism
$\Cur(t)\in\Kl\PCOH(Z,\Impl XY)$ defined as
$\Cur(t)=\Curlin(t\Compl\Expmonisobn)$. 

\subsection{Least fix-point operator in the Kleisli category.}
Let $X$ be an object of $\PCOH$. Let $\cF\in\Kl\PCOH(\Impl{(\Impl
  XX)}{X},\Impl{(\Impl XX)}{X})$ be $\cF=\Cur{(\cF_0)}$ where
$\cF_0\in\Kl\PCOH({(\Impl{(\Impl XX)}{X})}\IWith{(\Impl XX)},X)$ is the
following composition of morphisms in $\Kl\PCOH$
\begin{center}
  \begin{tikzpicture}[->, >=stealth]
    \node (1) {${(\Impl{(\Impl XX)}{X})}\IWith{(\Impl XX)}$};
    \node (2) [below of=1, node distance=1.2cm]
      {$(\Impl XX)\IWith{(\Impl{(\Impl XX)}{X})}\IWith{(\Impl XX)}$};
    \node (3) [right of=2, node distance=7cm]
      {$(\Impl XX)\IWith X$};
    \node (4) [above of=3, node distance=1.2cm] {$X$};
    \tikzstyle{every node}=[midway,auto,font=\scriptsize]
    \draw (1) -- node {$\Tuple{\Proj 2,\Proj 1,\Proj 2}$} (2);
    \draw (2) -- node {$\Tuple{\Proj 1,\Ev\Comp\Tuple{\Proj 2,\Proj 3}}$} (3);
    \draw (3) -- node [swap] {$\Ev$} (4);
  \end{tikzpicture}
\end{center}
Then, given $F\in\Pcoh{(\Impl{(\Impl XX)}{X})}$, that is $F\in\PCOH(\Impl
XX,X)$, one has $\Fun\cF(F)=\Ev\Comp\Tuple{\Id_{\Impl XX},F}\in\PCOH(\Impl
XX,X)$. Since $\cF$ is a morphism in $\PCOH$, the function $\Fun\cF$ is Scott
continuous and therefore has a least fix-point $\Sfix\in\PCOH(\Impl XX,X)$,
namely $\Sfix=\sup_{n\in\Nat}{\Fun\cF}^n(0)$ (the sequence
$({\Fun\cF}^n(0))_{n\in\Nat}$ is monotone in the cpo $\Pcoh{(\Impl{(\Impl
    XX)}X)}$ because $\Fun\cF$ is monotone).

If we set $\Sfix_n={\Fun\cF}^n(0)\in\PCOH(\Impl XX,X)$, we have $\Sfix_0=0$ and
$\Sfix_{n+1}=\Ev\Comp\Tuple{\Id,\Sfix_n}$ so that, given $f\in\PCOH(X,X)$, we
have $\Fun{\Sfix_n}(f)=\Fun f^n(0)$ and $\Fun\Sfix(f)=\sup_{n\in\Nat}{\Fun
  f}^n(0)$. So that $\Sfix$ is the usual least fix-point operator, and this
operation turns out to be a morphism in $\Kl\PCOH$, namely
$\Sfix\in\Kl\PCOH(\Impl XX,X)$. This means that this standard least fix-point
operator can be described as a power series, which is not completely obvious at
first sight.

\subsection{Coalgebras}\label{sec:coalgebras}
By definition, a coalgebra of the $\Excl\_$ comonad is a pair $(X,h)$ where $X$
is a PCS and $h\in\PCOH(X,\Excl X)$ satisfies the following commutations 

\begin{center}
  \begin{tikzpicture}[->, >=stealth]
    \node[matrix,matrix of math nodes, row sep=8mm] (D1)
    { |(X1)| X &[8mm] |(EX)| \Excl X\\
      & |(X2)| X\\
    };

    \node[matrix,matrix of math nodes, row sep=8mm, right of=D1, node distance=4cm] (D2)
    { |(X3)| X &[8mm] |(EX3)| \Excl X \\
      |(EX4)| \Excl X & |(EEX)| \Excl{\Excl X}\\
      };

    \tikzstyle{every node}=[midway,auto,font=\scriptsize]
    \draw (X1) -- node {$h$} (EX);
    \draw (EX) -- node {$\Der X$} (X2);
    \draw (X1) -- node [swap] {$\Id_X$}(X2);

    \draw (X3) -- node {$h$} (EX3);
    \draw (EX3) -- node {$\Excl h$} (EEX);
    \draw (X3) -- node[swap] {$h$} (EX4);
    \draw (EX4) -- node[swap] {$\Digg X$} (EEX);
  \end{tikzpicture}
\end{center}

A morphism from a coalgebra $(X_1,h_1)$ to a coalgebra $(X_2,h_2)$ is an
$f\in\PCOH(X_1,X_2)$ such that the following diagram commutes
\begin{center}
  \begin{tikzpicture}[->, >=stealth]
    \node[matrix,matrix of math nodes, row sep=8mm] (Dm)
    { |(X1)| X_1 &[8mm] |(X2)| X_2 \\
      |(EX1)| \Excl{X_1} &[8mm] |(EX2)| \Excl{X_2} \\
    };
    \tikzstyle{every node}=[midway,auto,font=\scriptsize]
    \draw (X1) -- node {$f$} (X2);
    \draw (EX1) -- node[swap] {$\Excl f$} (EX2);
    \draw (X1) -- node[swap] {$h_1$} (EX1);
    \draw (X2) -- node {$h_2$} (EX2);
  \end{tikzpicture}
\end{center}

Observe that $\One$ has a natural structure of $\oc$-coalgebra
$\upsilon\in\PCOH(\One,\Excl\One)$ which is obtained as the following
composition of morphisms
\begin{center}
  \begin{tikzpicture}
    [->, >=stealth]
    \node[matrix,matrix of math nodes, row sep=8mm]
    { |(1)| \One &[8mm] |(ET)| \Excl\Top
      &[8mm] |(EET)| \Excl{\Excl\Top} &[8mm] |(E1)| \Excl\One\\
    };
    \tikzstyle{every node}=[midway,auto,font=\scriptsize]
    \draw (1) -- node {$\Expmonisoz$} (ET);
    \draw (ET) node {$\Digg\Top$}-- (EET);
    \draw (EET) -- node {$\Excl{\Funinv{(\Expmonisoz)}}$} (E1);
  \end{tikzpicture}
\end{center}
Checking that $(\One,\upsilon)$ is indeed a $\oc$-coalgebra boils down to a
simple diagrammatic computation using the general axioms satisfied by the
comonadic and monoidal structure of the $\Excl\_$ functor.

A simple computation shows that $\upsilon_{*,n}=1$ for all
$n\in\Web{\Excl\One}$ (remember that $\Web{\Excl\One}=\Nat$). 
%
%

Let $(X_i,h_i)_{i\in I}$ be a countable family of coalgebras. Then we can endow
$X=\bigoplus_{i\in I}X_i$ with a structure of coalgebra $h\in\PCOH(X,\Excl
X)$. By the universal property of the coproduct, it suffices to define for each
$i\in I$ a morphism $h'_i:X_i\to\Excl X$. We set $h'_i=\Excl{\Inj i}\Compl h_i$
where we record that $\Inj i:X_i\to X$ is the $i$th canonical injection into
the coproduct. It is then quite easy to check that $(X,h)$ so defined is a
coalgebra using the fact that each $(X_i,h_i)$ is a coalgebra.

\paragraph{Natural numbers.}
Consider the case where $I=\Nat$, $X_i=\One$ and $h_i=\upsilon$ for each
$i\in\Nat$. Then we use $\Snat$ to denote the corresponding object $X$ and
$\Natalg$ for the corresponding coalgebra structure,
$\Natalg\in\PCOH(\Natobj,\Excl\Natobj)$. We use $\Snum n\in\PCOH(\One,\Natobj)$
for the $n$th injection that we consider also as the element of $\Pcoh\Natobj$
defined by $\Snum n_k=\Kronecker nk$. 

An easy computation shows that
\begin{align*}
  (\Natalg)_{n,\mu}=
  \begin{cases}
    1 & \text{if $\mu=k\Mset n$ for some $k\in\Nat$}\\
    0 & \text{otherwise.}
  \end{cases}
\end{align*}

Let $t\in\Kl\PCOH(\Natobj,X)$ for some object $X$ of $\PCOH$. Then
$t\Compl\Natalg\in\PCOH(\Natobj,X)$ is a linearized\footnote{This is not at all
  the same kind of linearization as the one introduced by Differential Linear
  Logic~\cite{Ehrhard15c}.} version of $t$. Given $u\in\Pcoh{\Natobj}$, an easy
computation shows that
\begin{align*}
  t\Compl\Natalg\Compl u=\sum_{n=0}^\infty u_n\Fun t(\Snum n)\,.
\end{align*}

The objects $\Natobj$ and $\Plus\One\Natobj$ are obviously isomorphic, through
the morphisms $p\in\PCOH(\Natobj,\Plus\One\Natobj)$ and
$s\in\PCOH(\Plus\One\Natobj,\Natobj)$ given by
\begin{align*}
  p_{n,(1,*)}=s_{(1,*),n}=\Kronecker n0\text{ and }
  p_{n,(2,n')}=s_{(2,n'),n}=\Kronecker {n}{n'+1}
\end{align*}
We set $\Ssuc=s\Compl\Inj 2\in\PCOH(\Natobj,\Natobj)$, so that
$\Ssuc_{n,n'}=\Kronecker{n+1}{n'}$ represents the successor function.

\subsection{Conditional}

Given an object $X$ of $\PCOH$, we define a morphism
\begin{align*}
 \Sif\in\PCOH(\Natobj\ITens\Excl X\ITens\Excl{(\Limpl{\Excl\Natobj}{X})},X)\,.
\end{align*}
For this, we define first 
$\Sif_0\in\PCOH(\One\ITens\Excl X\ITens\Excl{(\Limpl{\Excl\Natobj}{X})},X)$
as the following composition of morphisms (without mentioning the isomorphisms
associated with the monoidality of $\ITens$)
\begin{center}
  \begin{tikzpicture}[->, >=stealth]
    \node (1) [node distance=2cm]
       {$\Excl X\ITens\Excl{(\Limpl{\Excl\Natobj}{X})}$};
    \node (2) [right of=1, node distance=2.5cm] {$\Excl X$};
    \node (3) [right of=2, node distance=1.4cm] {$X$};
    \tikzstyle{every node}=[midway,auto,font=\scriptsize]
    \draw (1) -- node {$\Tens{\Excl X}{\Weak{}}$} (2);
    \draw (2) -- node {$\Der X$} (3);
  \end{tikzpicture}
\end{center}
and next $\Sif_{+}\in\PCOH(\Natobj\ITens\Excl
X\ITens\Excl{(\Limpl{\Excl\Natobj}{X})},X)$ (with the same conventions as
above)
\begin{center}
  \begin{tikzpicture}[->, >=stealth]
    \node (1) {$\Natobj\ITens\Excl X\ITens\Excl{(\Limpl{\Excl\Natobj}{X})}$};
    \node (2) [right of=1, node distance=4.8cm] 
      {$\Excl\Natobj\ITens{(\Limpl{\Excl\Natobj}{X})}$};
    \node (3) [right of=2, node distance=2cm] {$X$};
    \tikzstyle{every node}=[midway,auto,font=\scriptsize]
    \draw (1) -- node {$\Natalg\ITens\Weak{}\ITens\Der{}$} (2);
    \draw (2) -- node {$\Evlin\Compl\Sym$} (3);
  \end{tikzpicture}
\end{center}
where $\Sym$ is the isomorphism associated with the symmetry of the functor
$\ITens$, see Section~\ref{sec:multiplicatives}. 

The universal property of $\IPlus$ and the fact that $\Tens\_ Y$ is a left
adjoint for each object $Y$ allows therefore to define
$\Sif'\in\PCOH((\Plus\One\Natobj)\ITens\Excl
X\ITens\Excl{(\Limpl{\Excl\Natobj}X)},{X})$. Finally our conditional morphism
is $\Sif=\Sif'\Compl(p\ITens\Excl
X\ITens\Excl{(\Limpl{\Excl\Natobj}X)})\in\PCOH(\Natobj\ITens\Excl
X\ITens\Excl{(\Limpl{\Excl\Natobj}{X})},X)$. The isomorphism
$p\in\PCOH(\Natobj,\Plus\One\Natobj)$ is defined at the end of
Section~\ref{sec:coalgebras}.

It is important to notice that the two following diagrams commute
\begin{center}
  \begin{tikzpicture}[->, >=stealth]
    \node (1) {$\One\ITens\Excl X\ITens\Excl{(\Limpl{\Excl\Natobj}{X})}$}; 
    \node (2) [right of=1, node distance=4.4cm]
      {$\Natobj\ITens\Excl X\ITens\Excl{(\Limpl{\Excl\Natobj}{X})}$}; 
    \node (3) [below of=2, node distance=1.2cm] {$X$};
    \tikzstyle{every node}=[midway,auto,font=\scriptsize]
    \draw (1) -- node {$\Snum 0\ITens\Id$} (2);
    \draw (2) -- node {$\Sif$} (3);
    \draw (1) -- node [swap] {$\Tens{\Der{}}{\Weak{}}$} (3);
  \end{tikzpicture}
\end{center}

\begin{center}
  \begin{tikzpicture}[->, >=stealth]
    \node (1) {$\One\ITens\Excl X\ITens\Excl{(\Limpl{\Excl\Natobj}{X})}$}; 
    \node (2) [right of=1, node distance=4.6cm]
      {$\Natobj\ITens\Excl X\ITens\Excl{(\Limpl{\Excl\Natobj}{X})}$}; 
    \node (3) [below of=1, node distance=1.2cm] 
      {$\Excl\Natobj\ITens(\Limpl{\Excl\Natobj}{X})$};
    \node (4) [below of=2, node distance=1.2cm] {$X$};
    \tikzstyle{every node}=[midway,auto,font=\scriptsize]
    \draw (1) -- node {$\Snum{n+1}\ITens\Id$} (2);
    \draw (2) -- node {$\Sif$} (4);
    \draw (1) -- node [swap] {$\Tens{\Prom{\Snum n}}{\Weak{}}$} (3);
    \draw (3) -- (4) node {$\Evlin\Compl\Sym$};
  \end{tikzpicture}
\end{center}

This second commutation boils down to the following simple property: $\forall
n\in\Nat\ \Natalg\Compl\Snum n=\Prom{\Snum n}$. Observe that it is not true
however that $\forall u\in\Pcoh\Snat\ \Natalg\Compl u=\Prom{u}$. This means
that $\Natalg$ allows to duplicate and erase ``true'' natural numbers $\Snum n$
but not general elements of $\Pcoh\Snat$ which can be considered as
``computations'' and not as ``values''.

\subsection{Interpreting terms}

Given a type $\sigma$, we define an object $\Tsem\sigma$ of $\PCOH$ as follows:
$\Tsem\Tnat=\Natobj$ and
$\Tsem{\Timpl\sigma\tau}=\Impl{\Tsem\sigma}{\Tsem\tau}$. 

Given a context $\Gamma=(x_1:\sigma_1,\dots,x_k:\sigma_k)$, a type $\sigma$ and
a term $M$ such that $\Tseq\Gamma M\sigma$, we define a morphism $\Psem
M\Gamma\in\Kl\PCOH(\Tsem\Gamma,\Tsem \sigma)$ where
$\Tsem\Gamma=\Tsem{\sigma_1}\IWith\cdots\IWith\Tsem{\sigma_k}$. Equivalently,
we can see $\Psem M\Gamma$ as a morphism in $\PCOH(\Tseme\Gamma,\Tsem \sigma)$
where $\Tseme\Gamma=
\Excl{\Tsem{\sigma_1}\ITens\cdots\ITens\Excl{\Tsem{\sigma_k}}}$.
By Theorem~\ref{th:pcoh-functional}, this morphism can be fully described as a
function $\Fun{\Psem M\Gamma}:\prod_{i=1}^k\Pcoh{\Tsem{\sigma_i}}\to\Pcoh{\Tsem
  \sigma}$. The definition is by induction on the typing derivation of
$\Tseq\Gamma M\sigma$, or, equivalently, on $M$.

If $M=x_i$, then $\Psem M\Gamma=\Proj i$, that is $\Fun{\Psem
M\Gamma}(u_1,\dots,u_k)=u_i$. 

If $M=\Num n$, then $\Psem M\Gamma=\Snum n\Comp\tau$ where $\tau$ is the
unique morphism in $\Kl\PCOH(\Tsem\Gamma,\Top)$. That is $\Fun{\Psem
M\Gamma}(\Vect u)=\Snum n$.

If $M=\Dice p$ for some $p\in[0,1]\cap\Rational$ then $\Psem M\Gamma=p\Snum
0+(1-p)\Snum 1$.

If $M=\Succ P$ with $\Tseq\Gamma P\Tnat$, we have $\Psem
P\Gamma\in\PCOH(\Tseme\Gamma,\Natobj)$ and we set $\Psem
M\Gamma=\Ssuc\Compl\Psem P\Gamma$, which is characterized by $\Fun{\Psem
M\Gamma}(\Vect u)=\sum_{n=0}^\infty(\Fun{\Psem P\Gamma}(\Vect u))_n\Snum{n+1}$.

If $M=\Ifv PQzR$, $\Tseq\Gamma P\Tnat$, $\Tseq\Gamma Q\sigma$ and
$\Tseq{\Gamma,z:\Tnat}R\sigma$ then by inductive hypothesis we have $\Psem
P\Gamma\in\PCOH(\Tseme\Gamma,\Natobj)$, $\Psem
Q\Gamma\in\PCOH(\Tseme\Gamma,\Tsem \sigma)$ and $\Psem
R{\Gamma,z:\Tnat}\in\PCOH(\Tseme\Gamma\ITens\Excl\Natobj,\Tsem \sigma)$. We
have $\Curlin{(\Psem
  R{\Gamma,z:\Tnat})}\in\PCOH(\Tseme\Gamma,\Limpl{\Excl\Natobj}{\Tsem \sigma})$
and hence we define $\Psem M\Gamma$ as the following composition of morphisms
in $\PCOH$
\begin{center}
  \begin{tikzpicture}[->, >=stealth]
    \node (1) {$\Tseme\Gamma$};
    \node (2) [below of=1, node distance=1.2cm] 
      {$\Tseme\Gamma\ITens\Tseme\Gamma\ITens\Tseme\Gamma$};
    \node (3) [right of=2, node distance=6.6cm] 
      {$\Natobj\ITens\Excl{\Tsem \sigma}
               \ITens\Excl{(\Limpl{\Excl\Natobj}{\Tsem \sigma})}$};
    \node (4) [above of=3, node distance=1.2cm] {$\Tsem \sigma$}; 
    \tikzstyle{every node}=[midway,auto,font=\scriptsize]
    \draw (1) -- node {$\Contrc{\Gamma}$} (2);
    \draw (2) -- node {$\Psem M\Gamma\ITens\Prom{\Psem P\Gamma}
      \ITens\Prom{\Curlin{(\Psem R{\Gamma,z:\Tnat})}}$} (3);
    \draw (3) -- node {$\Sif$} (4);
  \end{tikzpicture}
\end{center}
where $\Contrc\Gamma$ is an obvious composition of contraction morphisms and
associativity and symmetry isomorphisms associated with the $\ITens$ functor
(we also use promotion~\Eqref{eq:promo-def}). Seen as a function, this morphism
is completely characterized by
\begin{align*}
  \Fun{\Psem M\Gamma}(\Vect u)=(\Fun{\Psem P\Gamma}(\Vect u))_0\Fun{\Psem
    Q\Gamma}(\Vect u) +\sum_{n=0}^\infty(\Fun{\Psem P{\Gamma}}(\Vect u))_{n+1}
  \Fun{\Psem R{\Gamma,z:\Tnat}}(\Vect u,\Snum n)\,.
\end{align*}

If $M=\App PQ$ with $\Tseq\Gamma P{\Timpl \sigma\tau}$ and $\Tseq\Gamma
Q\sigma$ then we have $\Psem P\Gamma\in\Kl\PCOH(\Tseme\Gamma,\Limpl{\Excl{\Tsem
    \sigma}}{\Tsem \tau})$ and $\Psem Q\Gamma\in\Kl\PCOH(\Tseme\Gamma,\Tsem
\sigma)$ and we define $\Psem M\Gamma$ as the following composition of
morphisms
\begin{center}
  \begin{tikzpicture}[->, >=stealth]
    \node (1) {$\Tseme\Gamma$};
    \node (2) [right of=1, node distance=2.2cm] 
      {$\Tseme\Gamma\ITens\Tseme\Gamma$};
    \node (3) [right of=2, node distance=4.2cm] 
      {$(\Limpl{\Excl{\Tsem \sigma}}{\Tsem \tau})
               \ITens\Excl{\Tsem \sigma}$};
    \node (4) [right of=3, node distance=2.4cm] {$\Tsem \tau$}; 
    \tikzstyle{every node}=[midway,auto,font=\scriptsize]
    \draw (1) -- node {$\Contrc{\Gamma}$} (2);
    \draw (2) -- node {$\Psem P\Gamma\ITens\Prom{\Psem Q\Gamma}$} (3);
    \draw (3) -- node {$\Evlin$} (4);
  \end{tikzpicture}
\end{center}
so that $\Psem M\Gamma$ is characterized by $\Fun{\Psem M\Gamma}(\Vect
u)=\Fun{\Psem P\Gamma}(\Vect u)(\Fun{\Psem Q\Gamma}(\Vect u))$.

If $M=\Abst x\sigma P$ with $\Tseq{\Gamma,x:\sigma}P\tau$ then we have
$\Psem{P}{\Gamma,x:\sigma}\in\PCOH(\Tseme\Gamma\ITens\Excl{\Tsem \sigma},\Tsem
\tau)$ and we set $\Psem{M}\Gamma=\Curlin{(\Psem{P}{\Gamma,x:\sigma})}
\in\PCOH(\Tseme\Gamma,\Limpl{\Excl{\Tsem \sigma}}{\Tsem \tau})$ so that, given
$\Vect u\in\prod_{i=1}^k\Pcoh{\Tsem{\sigma_i}}$ (remember that
$\Gamma=(x_1:\sigma_1,\dots,x_k:\sigma_k)$), the semantics $\Psem M\Gamma(\Vect
u)$ of $M$ is the element of $\Pcoh{(\Limpl{\Excl{\Tsem \sigma}}{\Tsem \tau})}$
which, as a function $\Pcoh{\Tsem \sigma}\to\Pcoh{\Tsem \tau}$, is
characterized by $\Fun{\Psem M\Gamma}(\Vect u)(u)=\Fun{\Psem
  P{\Gamma,x:\sigma}}(\Vect u,u)$.

If $M=\Fix P$ with $\Tseq\Gamma P{\Timpl \sigma\sigma}$ then we have $\Psem
P\Gamma\in\PCOH(\Tseme\Gamma,\Limpl{\Excl{\Tsem\sigma}}{\Tsem\sigma})$ and we
set $\Psem M\Gamma=\Sfix\Compl\Prom{\Psem P\Gamma}$. This means that
$\Fun{\Psem M\Gamma}(\Vect u)=\sup_{n\in\Nat}f^n(0)$ where $f\in\Kl\PCOH(\Tsem
\sigma,\Tsem \sigma)$ is given by $f(u)=\Fun{\Psem P\Gamma}(\Vect u)(u)$.

\begin{lemma}[Substitution]
  Assume that $\Tseq{\Gamma,x:\sigma}{M}{\tau}$ and that $\Tseq\Gamma
  P\sigma$. Then $\Psem{\Subst MPx}\Gamma=\Psem
  M{\Gamma,x:\sigma}\Comp\Tuple{\Id_{\Tsem\Gamma},\Psem P\Gamma}$ in
  $\Kl\PCOH$. In other words, for any $\Vect u\in\Pcoh{\Tsem\Gamma}$, we have
  $\Fun{\Psem{\Subst MPx}\Gamma}(\Vect u)=\Fun{\Psem M{\Gamma,x:\sigma}}(\Vect
  u,\Fun{\Psem P\Gamma}(\Vect u))$.
\end{lemma}
The proof is a simple induction on $M$, the simplest way to write it is to use
the functional characterization of the semantics.

For the notations $\Open\Gamma\sigma$ and $\Closed\sigma$ used below, we refer
to \Parag{par:stoch-mat-red}. We formulate the invariance of the interpretation
of terms under weak-reduction, using the stochastic reduction matrix introduced
in \Parag{par:stoch-mat-red}.
\begin{theorem}\label{th:sem-invar}
  Assume that $\Tseq\Gamma M\sigma$. One has
  \begin{align*}
    \Psem M\Gamma=\sum_{M'\in\Open\Gamma\sigma}
      \Redmato\Gamma\sigma_{M,M'}\Psem{M'}\Gamma
  \end{align*}
\end{theorem}
\Beginproof
Simple case analysis, on the shape of $M$, and using the Substitution Lemma.
\Endproof

As a corollary we get the following inequality.
\begin{theorem}\label{th:soundness-ineq}
  Let $M$ be such that $\Tseq{}M\Tnat$. Then for all $n\in\Nat$ we have
  \begin{align*}
    \Redmat\Tnat^\infty_{M,\Num n}\leq{\Psem M{}}_n\,.
  \end{align*}
\end{theorem}
\Beginproof
Iterating Theorem~\ref{th:sem-invar} we get, for all $k\in\Nat$:
\begin{align*}
  \Psem M{}=\sum_{M'\in\Closed\Tnat}\Redmat\Tnat^k_{M,M'}\Psem{M'}{}
\end{align*}
Therefore, for all $k\in\Nat$ we have ${\Psem M{}}_n\geq\Redmat\Tnat^k_{M,\Num
  n}$ and the result follows, since $\Num n$ is weak-normal.
\Endproof

\paragraph{Examples.}\label{sec:interp-examples}
We refer to the various terms introduced in Section~\ref{sec:basic-examples}
and describe as functions the interpretation of some of them.

We have $\Tseq{}{\Pred}{\Timpl\Tnat\Tnat}$ so
$\Psem\Pred{}\in\Pcoh{(\Impl\Snat\Snat)}$, and one checks easily that
$\Fun{\Psem\Pred{}}(u)=(u_0+u_1)\Snum 0+\sum_{n=1}^\infty u_{n+1}\Snum n$.

Similarly, we have
\begin{align*}
  \Fun{\Psem\Add{}}(u)(v)
  &=\sum_{n=0}^\infty\Big(\sum_{i=0}^n u_iv_{n-i}\Big)\Snum n\\
  \Fun{\Psem\Exp{}}(u)&=\sum_{n=0}^\infty u_n\Snum{2^n}\\
  \Fun{\Psem{\Shift_k}{}}(u)&=\sum_{n=0}^\infty u_n\Snum{k+n}\\
  \Fun{\Psem\Cmp{}}(u)(v)&=\Big(\sum_{i\leq j}u_iv_j\Big)\Snum 0
  +\Big(\sum_{i>j}u_iv_j\Big)\Snum 1\\
  \Fun{\Psem{\Probe_k}{}}(u)&=u_k\Snum 0\\
  \Fun{\Psem{\Pprod_k}{}}(u^1,\dots,u^k)&=
  \Big(\prod_{i=1}^ku^i_0\Big)\Snum 0\\
  \Fun{\Psem{\Pchoose_k}{}}(u)(w^1,\dots,w^k)&=
  \sum_{i=0}^{k-1}u_iw^{i+1}\\
  \Fun{\Psem{\Unif}{}}(u)&=
  \sum_{n=0}^\infty\frac{u_n}{n+1}\Big(\sum_{i=0}^n\Snum i\Big)
  =\sum_{i=0}^\infty\Big(\sum_{n=i}^\infty\frac{u_n}{n+1}\Big)\Snum i\\
  \Psem{\Ran{\List p0n}}{}&=\sum_{i=0}^np_i\Snum i
\end{align*}

\section{Adequacy}
We want now to prove the converse inequality to that of
Theorem~\ref{th:soundness-ineq}.


For any type $\sigma$ we define a binary relation
$\Rts\sigma\subseteq\Closed\sigma\times\Pcoh{\Tsem\sigma}$ by induction on
types as follows:
\begin{itemize}
\item $M\Rel{\Rts\Tnat}u$ if
  $\forall n\in\Nat\ u_n\leq\Redmat\Tnat^\infty_{M,\Num n}$
\item $M\Rel{\Rts{\Timpl\sigma\tau}}t$ if $\forall P\in\Closed\sigma\,\forall
  u\in\Pcoh{\Tsem\sigma}\ P\Rel{\Rts\sigma}u\Implies \App MP\Rel{\Rts\tau}\Fun
  t(u)$\,. Here we have $t\in\Pcoh{\Tsem{\Timpl\sigma\tau}}$ and hence $\Fun
  t:\Pcoh{\Tsem\sigma}\to\Pcoh{\Tsem\tau}$
\end{itemize}
So $\Rts\sigma$ is a logical relation.

\begin{lemma}\label{lemma:rts-order}
  If $M\in\Closed\sigma$ then $M\Rel{\Rts\sigma}0$. If $(u(i))_{i\in\Nat}$ is an
  increasing sequence in $\Pcoh{\Tsem\sigma}$ such that $\forall i\in\Nat\
  M\Rel{\Rts\sigma}u(i)$, then $M\Rel{\Rts\sigma}\sup_{i\in\Nat}u(i)$.
\end{lemma}
\Beginproof
Simple induction on types, using Proposition~\ref{prop:order-fun-kleiseli}.
\Endproof

\begin{lemma}\label{lemma:ifv-red-sem}
  Assume that $\Tseq{}M\Tnat$, $\Tseq{}P\sigma$ and $\Tseq{z:\Tnat}Q\sigma$
  where $\sigma=\Timpl{\sigma_1}{\cdots\Timpl{\sigma_k}{\Tnat}}$. Let $\List
  N1k$ be terms such that $\Tseq{}{N_i}{\sigma_i}$ for $i=1,\dots,k$.

  Then, for any $n\in\Nat$, we have
  \begin{align*}
    &\Redmat\Tnat^\infty_{\App{\Ifv MPzQ}{N_1\cdots N_k},\Num n}\\
    &\quad=
    \Redmat\Tnat^\infty_{M,\Num 0}
    \Redmat\Tnat^\infty_{\App P{N_1\cdots N_k},\Num n}
    +\sum_{k\in\Nat}
       \Redmat\Tnat^\infty_{M,\Num{k+1}}
       \Redmat\Tnat^\infty_{\App{\Subst Q{\Num k}z}{N_1\cdots N_k},\Num n}
  \end{align*}
\end{lemma}
This is a straightforward consequence of the definition of weak-reduction,
and of Lemma~\ref{lemma:stochinf-paths}.

\begin{lemma}\label{lemma:rts-red-closed}
  Let $\sigma$ be a type.  Let $M,M'\in\Closed\sigma$ and let
  $u\in\Pcoh{\Tsem\sigma}$. Then
  \begin{align*}
  M'\Rel{\Rts\sigma}u\Implies M\Rel{\Rts\sigma}\Redmat\sigma_{M,M'}u\,.
  \end{align*}
\end{lemma}
\Beginproof
By induction on $\sigma$. Assume first that $\sigma=\Tnat$.

Assume that $M'\Rel{\Rts\Tnat}u$. This means that, for all $n\in\Nat$, one has
$u_n\leq\Redmat\Tnat^\infty_{M',\Num n}$. Let $n\in\Nat$, we want to prove that
\begin{align*}
  \Redmat\Tnat^\infty_{M,\Num n}\geq\Redmat\Tnat_{M,M'}u_n
\end{align*}
This results from the fact that $\Redmat\Tnat^\infty_{M,\Num
  n}=\sum_{M''\in\Closed\Tnat}\Redmat\Tnat_{M,M''}\Redmat\Tnat^\infty_{M'',\Num
  n}$ and from our hypothesis about $M'$.

Assume now that $\sigma=\Timpl\tau\phi$ and let
$f\in\Pcoh{\Tsem\sigma}$. Assume that $M'\Rel{\Rts{\Timpl\tau\phi}}f$, we want
to prove that
\begin{align*}
  M\Rel{\Rts{\Timpl\tau\phi}}\Redmat{\Timpl\tau\phi}_{M,M'}f
\end{align*}
If $M$ is weak-normal then either $M'=M$ and then
$\Redmat{\Timpl\tau\phi}_{M,M'}=1$ and we can directly apply our hypothesis
that $M'\Rel{\Rts{\Timpl\tau\phi}}f$, or $M'\not=M$ and then
$\Redmat{\Timpl\tau\phi}_{M,M'}=0$, and we can apply
Lemma~\ref{lemma:rts-order}. So assume that $M$ is not weak-normal.

Let $P\in\Closed\tau$ and $u\in\Pcoh{\Tsem\tau}$ be such that
$P\Rel{\Rts\tau} u$. We need to prove that 
\begin{align*}
  \App MP\Rel{\Rts{\phi}}\Redmat{\phi}_{M,M'}f(u)\,.
\end{align*}

This results from the inductive hypothesis and from the fact that, due to our
definition of weak-reduction, it holds that
$\Redmat{\Timpl\tau\phi}_{M,M'}=\Redmat{\phi}_{\App MP,\App{M'}P}$ because $M$
is not weak-normal.
\Endproof

\begin{remark}
  From now on, and for the purpose of avoiding too heavy notations, we often
  consider implicitly morphisms of $\Kl\PCOH$ as functions. Typically, if
  $f\in\Kl\PCOH(X,Y)$ and $u\in\Pcoh X$, we write as above $f(u)$
  instead of $\Fun f(u)$.
\end{remark}

\begin{theorem}
  Assume that $\Tseq{\Gamma}{M}{\sigma}$ where
  $\Gamma=(x_1:\sigma_1,\dots,x_l:\sigma_l)$. For all families $(P_i)_{i=1}^l$
  and $(u_i)_{i=1}^l$ one has
  \begin{align*}
    (\forall i\ P_i\Rel{\Rts{\sigma_i}}u_i)
    \Implies \Substbis M{P_1/x_1,\dots,P_l/x_l}
    \Rel{\Rts\sigma}\Psem M\Gamma(u_1,\dots,u_l)
  \end{align*}
\end{theorem}
\Beginproof
By induction on the derivation of $\Tseq{\Gamma}{M}{\sigma}$ (that is, on $M$).

\Proofbreak
The cases $M=x_i$ and $M=\Num n$ are straightforward.

\Proofbreak
Assume that $M=\Dice p$ where $p\in[0,1]\cap\Rational$. Then $\sigma=\Tnat$ and
$\Psem M\Gamma(\List u1l)=p\Snum 0+(1-p)\Snum 1$. On the other hand
\begin{align*}
  \Redmat\Tnat_{\Substbis M{P_1/x_1,\dots,P_l/x_l},\Num n}
  =
  \begin{cases}
    p & \text{if } n=0\\
    1-p & \text{if } n=1\\
    0 & \text{otherwise}
  \end{cases}
\end{align*}
and hence $(\forall i\ P_i\Rel{\Rts{\sigma_i}}u_i) \Implies \Substbis
M{P_1/x_1,\dots,P_l/x_l} \Rel{\Rts\sigma}\Psem M\Gamma(u_1,\dots,u_l)$ by
definition of $\Rts\Tnat$.

\Proofbreak
Assume that $M=\Succ N$. Assume that $\forall i\
P_i\Rel{\Rts{\sigma_i}}u_i$. By inductive hypothesis we have $\Substbis
N{P_1/x_1,\dots,P_l/x_l} \Rel{\Rts\Tnat}\Psem N\Gamma(u_1,\dots,u_l)$. This
means that, for all $n\in\Nat$, one has
\begin{align*}
  \Psem N\Gamma(u_1,\dots,u_l)_n\leq
  \Redmat\Tnat^\infty_{\Substbis N{P_1/x_1,\dots,P_l/x_l},\Num
  n}
\end{align*}
It follows that, for all $n\in\Nat$,
\begin{align*}
  \Psem{\Succ N}\Gamma(u_1,\dots,u_l)_{n+1}\leq
  \Redmat\Tnat^\infty_{\Substbis{\Succ N}{P_1/x_1,\dots,P_l/x_l},\Num{n+1}}
\end{align*}
that is
\begin{align*}
  \forall n\in\Nat\quad\Psem{\Succ N}\Gamma(u_1,\dots,u_l)_{n}\leq
  \Redmat\Tnat^\infty_{\Substbis{\Succ N}{P_1/x_1,\dots,P_l/x_l},\Num{n}}
\end{align*}
since the inequality is obvious for $n=0$.

\Proofbreak
Assume that $M=\Ifv PQzR$ with $\Tseq\Gamma P\Tnat$, $\Tseq\Gamma Q\sigma$ and
$\Tseq{\Gamma,z:\Tnat} R\sigma$ with
$\sigma=\Timpl{\tau_1}{\Timpl{\cdots\tau_h}{\Tnat}}$. Assume that $\forall
i\ P_i\Rel{\Rts{\sigma_i}}u_i$. By inductive hypothesis, applying the
definition of $\Redmat\Tnat$, we get
\begin{align}
  \forall n\in\Nat\quad
  \Redmat\Tnat^\infty_{\Substbis P{P_1/x_1,\dots,P_l/x_l},\Num n}
  &\geq\Psem P\Gamma(\List u1l)_n\label{eq:hr-if-1}\\
  \Substbis Q{P_1/x_1,\dots,P_l/x_l}
  &\Rel{\Rts\sigma} \Psem Q\Gamma(\List u1l)\label{eq:hr-if-2}\\
  \forall n\in\Nat\quad
  \Substbis R{P_1/x_1,\dots,P_l/x_l,\Num n/z}
  &\Rel{\Rts\sigma}\Psem R{\Gamma,z:\Tnat}(\List u1l,\Snum n)
  \label{eq:hr-if-3}
\end{align}
Observe that, in the last equation, we use the inductive hypothesis with $l+1$
parameters, and we use the fact that, obviously, $\Num k\Rel{\Rts\Tnat}\Snum
k$. On the other hand, we have
\begin{align*}
  \Psem{M}\Gamma(\List u1l)&=
  \Psem P\Gamma(\List u1l)_0\Psem Q\Gamma(\List u1l)\\
  &\quad\quad+\sum_{k=0}^\infty
     \Psem P\Gamma(\List u1l)_{k+1}\Psem R{\Gamma,z:\Tnat}(\List
  u1l,\Snum k)
\end{align*}
and we must prove that $\Substbis
M{P_1/x_1,\dots,P_l/x_l}\Rel{\Rts\sigma}\Psem{M}\Gamma(\List u1l)$. So, for
$j=1,\dots,h$, let $R_j$ and $v_j$ be such that $\Tseq{}{R_j}{\tau_j}$,
$v_j\in\Pcoh{\Tsem{\tau_j}}$ and $R_j\Rel{\Rts{\tau_j}}v_j$. We must prove that
$\App{\Substbis M{P_1/x_1,\dots,P_l/x_l}}{R_1\cdots R_h}\Rel{\Rts\Tnat}\Psem
M\Gamma(\List u1l)(v_1)\cdots(v_h)$. Let $n\in\Nat$. By
Lemma~\ref{lemma:ifv-red-sem} we have
\begin{align*}
  &\Redmat\Tnat^\infty_{\App{\Substbis{M}{P_1/x_1,\dots,P_l/x_l}}
    {R_1\cdots R_h},\Num n}\\
  &\quad= \Redmat\Tnat^\infty_{\Substbis{P}{P_1/x_1,\dots,P_l/x_l},\Num 0}
  \Redmat\Tnat^\infty_{\App{\Substbis Q{P_1/x_1,\dots,P_l/x_l}}
    {R_1\cdots R_h},\Num n}\\
  &\quad\quad+\sum_{k=0}^\infty
  \Redmat\Tnat^\infty_{\Substbis{P}{P_1/x_1,\dots,P_l/x_l},\Num{k+1}}
  \Redmat\Tnat^\infty_{\App{\Substbis R{P_1/x_1,\dots,P_l/x_l,\Num k/z}}
    {R_1\cdots R_h},\Num n}
\end{align*}
By~\Eqref{eq:hr-if-1},~\Eqref{eq:hr-if-2} and~\Eqref{eq:hr-if-3}, and by
definition of $\Rts\sigma$, we have therefore
\begin{align*}
  &\Redmat\Tnat^\infty_{\App{\Substbis{M}{P_1/x_1,\dots,P_l/x_l}}
    {R_1\cdots R_h},\Num n}\\
  &\quad\geq \Psem{P}\Gamma(\List u1l)_0
  \Psem Q\Gamma(\List u1l)(v_1)\cdots(v_h)_{\Num n}\\
  &\quad\quad\quad+\sum_{k=0}^\infty
  \Psem P\Gamma(\List u1l)_{\Num{k+1}}
  \Psem R\Gamma(\List u1l,\Num k/z)(v_1)\cdots(v_h)_{\Num n}\\
  &\quad\quad=\Psem M\Gamma(\List u1l)(v_1)\cdots(v_h)_{\Num n}\\
\end{align*}
that is $\App{\Substbis M{P_1/x_1,\dots,P_l/x_l}}{R_1\cdots
  R_h}\Rel{\Rts\Tnat}\Psem M\Gamma(\List u1l)(v_1)\cdots(v_h)$ as contended.

\Proofbreak Assume that $M=\App PQ$ with $\Tseq\Gamma P{\Timpl\tau\sigma}$ and
$\Tseq\Gamma Q{\tau}$. Let $t=\Psem P\Gamma(\List u1l)$. Assume that $\forall
i\ P_i\Rel{\Rts{\sigma_i}}u_i$. By inductive hypothesis we have
\begin{align*}
  \Substbis P{P_1/x_1,\dots,P_l/x_l}\Rel{\Rts{\Timpl\tau\sigma}}t
\end{align*}
and $\Substbis Q{P_1/x_1,\dots,P_l/x_l}\Rel{\Rts{\tau}}\Psem Q\Gamma(\List
u1l)$. Hence we have
\begin{align*}
\Substbis{(\App PQ)}{P_1/x_1,\dots,P_l/x_l}\Rel{\Rts{\tau}}
    \Fun t(\Psem Q\Gamma(\List
u1l))
\end{align*}
which is the required property since $\Fun t(\Psem Q\Gamma(\List
u1l))=\Psem{\App PQ}{\Gamma}(\List u1l)$ by definition of the interpretation of
terms.

\Proofbreak Assume that $\sigma=(\Timpl\tau\phi)$, $M=\Abst x\tau P$ with
$\Tseq{\Gamma,x:\tau}P\phi$. Let $t=\Psem{\Abst x\tau P}\Gamma(\List
u1l)$. Assume also that $\forall i\ P_i\Rel{\Rts{\sigma_i}}u_i$. We must prove
that
\begin{align*}
  \Abst x\tau{(\Substbis P{P_1/x_1,\dots,P_l/x_l})}
  \Rel{\Rts{\Timpl\tau\phi}}t\,.
\end{align*}
To this end, let $Q$ be such that $\Tseq{}Q\tau$ and $v\in\Pcoh{\Tsem\tau}$
be such that $Q\Rel{\Rts\tau}v$, we have to make sure that
\begin{align*}
  \App{\Abst x\tau{(\Substbis P{P_1/x_1,\dots,P_l/x_l})}}Q
  \Rel{\Rts{\phi}}
  \Fun t(v)\,.
\end{align*}
By Lemma~\ref{lemma:rts-red-closed}, it suffices to prove that $\Substbis
P{P_1/x_1,\dots,P_l/x_l,Q/x} \Rel{\Rts{\phi}} \Fun t(v)$. This results from the
inductive hypothesis since we have $\Fun t(v)=\Psem P{\Gamma,x:\tau}(\List
u1n,v)$ by cartesian closeness.

\Proofbreak
Last assume that $M=\Fix P$ with $\Tseq\Gamma P{\Timpl\sigma\sigma}$. Assume
also that $\forall i\ P_i\Rel{\Rts{\sigma_i}}u_i$. We must prove that
\begin{align*}
  \Fix{\Substbis P{P_1/x_1,\dots,P_l/x_l}}
  \Rel{\Rts{\sigma}}
  \Psem{\Fix P}\Gamma(\List u1l)=\sup_{k=0}^\infty \Fun t^k(0)
\end{align*}
where $t=\Psem P\Gamma(\List
u1l)\in\Pcoh{(\Simpl{\Tsem\sigma}{\Tsem\sigma})}$. By
Lemma~\ref{lemma:rts-order}, it suffices to prove that
\begin{align*}
  \forall k\in\Nat\quad
  \Fix{\Substbis P{P_1/x_1,\dots,P_l/x_l}}
  \Rel{\Rts{\sigma}}\Fun t^k(0)
\end{align*}
and we proceed by induction on $k$. The base case $k=0$ results from
Lemma~\ref{lemma:rts-order}. Assume now that $\Fix{\Substbis
  P{P_1/x_1,\dots,P_l/x_l}} \Rel{\Rts{\sigma}}\Fun t^k(0)$ and let us prove
that
\begin{align*}
  \Fix{\Substbis P{P_1/x_1,\dots,P_l/x_l}}
  \Rel{\Rts{\sigma}}\Fun t^{k+1}(0)\,.
\end{align*}
By Lemma~\ref{lemma:rts-red-closed}, it suffices to prove that
\begin{align*}
  \App{\Substbis P{P_1/x_1,\dots,P_l/x_l}}
    {\Fix{\Substbis P{P_1/x_1,\dots,P_l/x_l}}}
  \Rel{\Rts{\sigma}}\Fun t^{k+1}(0)=\Fun t(\Fun t^{k}(0))\,.
\end{align*}
which results from the ``internal'' inductive hypothesis 
\[
\Fix{\Substbis P{P_1/x_1,\dots,P_l/x_l}} \Rel{\Rts{\sigma}}\Fun t^{k}(0)
\]
and from the ``external'' inductive hypothesis 
\[
\Substbis P{P_1/x_1,\dots,P_l/x_l}\Rel{\Rts{\Timpl\sigma\sigma}}t\,.
\]
\Endproof

In particular, if $\Tseq{}M\Tnat$ we have $\forall n\in\Nat\
\Redmat\Tnat^\infty_{M,\Num n}\geq(\Psem M{})_n$. By
Theorem~\ref{th:soundness-ineq} we have therefore the following
operational interpretation of the semantics of ground type closed terms.

\begin{theorem}\label{th:Nat-sem-proba}
  If $\Tseq{}M\Tnat$ then, for all $n\in\Nat$ we have $\forall n\in\Nat\
\Redmat\Tnat^\infty_{M,\Num n}=(\Psem M{})_n$.
\end{theorem}

As usual, the Adequacy Theorem follows straightforwardly. The observational
equivalence relation on terms is defined in Section~\ref{sec:obseq}.

\begin{lemma}\label{lemma:sem-modular}
  Given an observation context $\Thole C{\Gamma}\Tnat$, there is a function
  $f_C$ such that, for any term $M\in\Open\Gamma\sigma$, one has
  $\Psem{\Thsubst CM}=f_C(\Psem M\Gamma)$.
\end{lemma}
The proof is a simple induction on $C$.

\begin{theorem}[Adequacy]\label{th:eq-adequacy}
  Let $M,M'\in\Open\Gamma\sigma$ be terms of $\PCFP$. If $\Psem
  M{\Gamma}=\Psem{M'}{\Gamma}$ then $M\Rel\Obseq M'$.
\end{theorem}
\Beginproof
Assume that $\Psem M{\Gamma}=\Psem{M'}{\Gamma}$.  Let $\Thole C\Gamma\sigma$ be
an observation context such that $\Tseqh{}{C}\Gamma\sigma{\Tnat}$, we have
\begin{align*}
  \Redmat\Tnat^\infty_{\Thsubst CM,\Num 0}
  &= {\Psem{\Thsubst CM}{}}_0\text{\quad by Theorem~\ref{th:Nat-sem-proba}}\\
  &= f_C(\Psem M{\Gamma})_0\text{\quad by Lemma~\ref{lemma:sem-modular}}\\
  &= f_C(\Psem{M'}{\Gamma})_0\\
  &= \Redmat\Tnat^\infty_{\Thsubst C{M'},\Num 0}\,.
\end{align*}
\Endproof

\section{Full abstraction}\label{sec:fullabs}
We want now to prove the converse of Theorem~\ref{th:eq-adequacy}, that is:
given two terms $M$ and $M'$ such that $\Tseq{\Gamma}{M}{\sigma}$ and
$\Tseq{\Gamma}{M'}{\sigma}$, if $M\Rel{\Obseq}M'$ then
$\Psem{M}{\Gamma}=\Psem{M'}{\Gamma}$. This means that $\PCOH$ provides an
equationally fully abstract model of $\PPCF$.

\subsection{Intuition}
Let us first convey some intuitions about our approach to Full Abstraction. The
first thing to say is that the usual method, which consists in proving that the
model contains a collection of definable elements which is ``dense'' in a
topological sense, does not apply here because definable elements are very
sparse in $\PCOH$. 
%
%
For instance, in $\Pcoh{\Tsem{\Timpl\Tnat\Tnat}}$, there is an element $t$
which is characterized by $\Fun t(u)=4u_0u_1\Snum 0$. We have
$t\in\Pcoh{\Tsem{\Timpl\Tnat\Tnat}}$ because, for any $u\in\Pcoh\Snat$ we have
$u_0+u_1\leq 1$ and hence $u_0u_1\leq u_0(1-u_0)\leq 1/4$, and therefore $\Fun
t(u)\in[0,1]$. It is easy to see that $t$ is not definable in $\PPCF$. The
``best'' definable approximation of $t$ is obtained by means of the term
$\Abst{x}{\Tnat} {\Ifv x {\Ifv x {\Loopt\Tnat} {z'} {\Ifv{z'} {\Num 0} {z''}
      {\Loopt\Tnat} } } z {\Loopt\Tnat} } $ whose interpretation $s$ satisfies
$\Fun s(u)=2u_0u_1\Snum 0$.

Let $M$ and $M'$ be terms (that we suppose closed for simplifying and without
loss of generality) such that $\Tseq{}{M}{\sigma}$ and
$\Tseq{}{M'}{\sigma}$. Assume that $\Psem{M}{}\not=\Psem{M'}{}$, we have to
prove that $M\Rel{\not\Obseq}M'$. Let $a\in\Web{\Tsem\sigma}$ be such that
${\Psem{M}{}}_a\not={\Psem{M'}{}}_a$. We define a term $F$ such that
$\Tseq{}{F}{\Timpl\sigma\Tnat}$ and ${\Psem{\App F{M}}{}}_0\not={\Psem{\App
    F{M'}}{}}_0$. Then we use the observation context $C=\App F{\Hole{}\sigma}$
to separate $M$ and $M'$. For defining $F$, \emph{independently of $M$ and
  $M'$}, we associate with $a$ a closed term $\Ntest a$ such that
$\Tseq{}{\Ntest a}{\Timpl{\Tnat}{\Timpl{\sigma}{\Tnat}}}$ and which has the
following essential property:

\begin{quote}
  There is an $n\in\Nat$ --~depending only on $a$~-- such that, given
  $w,w'\in\Pcoh{\Tsem\sigma}$ such that $w_a\not=w'_a$ there are rational
  numbers $\List p0{n-1}\in\Rseg 01$ such that $\Fun{\Psem{\Ntest
      a}{}}(u)(w)_0\not=\Fun{\Psem{\Ntest a}{}}(u)(w')_0$ where $u=p_0\Snum
  0+\cdots+p_{n-1}\Snum{n-1}$.
\end{quote}

Applying this property to $w=\Psem M{}$ and $w'=\Psem{M'}{}$, we obtain the
required term $F$ by setting $F=\App{\Ntest a}{\Ran{\List p0{n-1}}}$.

In order to prove this crucial property of $\Ntest a$, we consider the map
$\phi_w:u\mapsto\Fun{\Psem{\Ntest a}{}}(u)(w)_0$ which is an entire function
depending only on the $n$ first components $u_0,\dots,u_{n-1}$ of
$u\in\Pcoh{\Snat}$ (again, $n$ is a non-negative integer which depends only on
$a$). 

\begin{quote}
  In Lemma~\ref{lemma:FA-one-coef}, we prove that the coefficient in $\phi_w$
  of the particular monomial $u_0u_1\dots u_{n-1}$ is $w_a$.
\end{quote}

 It follows that the functions $\phi_w$ and
$\phi_{w'}$ are different, and therefore take different values on an argument
of shape $p_0\Snum 0+\cdots+p_{n-1}\Snum{n-1}$ where all $p_i$s are rational,
because $\phi_w$ and $\phi_{w'}$ are continuous functions.

\subsection{Useful notions and constructs}\label{sec:basic-notions}
%
%
We introduce some elementary material used in the proof.
\begin{itemize}
\item First, for a morphism $t\in\Kl\PCOH(\Snat,X)$, we explain what it means
  to depend on finitely many parameters, considering $t$ as a function from a
  subset of $\Realpto\Nat$ to $\Pcoh X$.
\item Then we give the construction of the term $\Ntest a$ (testing term)
  and of the auxiliary term $\Ptest a$. The interpretations of these terms are
  morphisms depending on a finite number of parameters; we define explicitly
  $\Nlen a,\Plen a\in\Nat$ which are the number of relevant parameters. We also
  give the interpretation of these morphisms as functions in the category
  $\Kl\PCOH$.
\item We introduce next useful notations which will be used in the proof of the
  main lemma.
\end{itemize}
%
%

\renewcommand\Fun[1]{#1}

\paragraph{Morphisms depending on a finite number of parameters.}
Let $k\in\Nat$. Let $\Simplex k=\overbrace{\One\IPlus\cdots\IPlus\One}^k$ so
that $\Web{\Simplex k}=\{0,\dots,k-1\}$ and $\Pcoh{\Simplex k}=\{x\in\Realpto
k\St x_0+\dots+x_{k-1}\leq 1\}$. We have two morphisms
$\Embi(k)\in\Kl\PCOH(\Simplex k,\Pnat)$ and $\Embp(k)\in\Kl\PCOH(\Pnat,\Simplex
k)$ defined by
\begin{align*}
  \Embi(k)_{m,j}=\Embp(k)_{m,j}=
  \begin{cases}
    1 & \text{if }m=\Mset j\text{ and }j<k\\
    0 & \text{otherwise.}
  \end{cases}
\end{align*}
Given $t\in\Kl\PCOH(\Pnat,X)$, the morphism
$s=t\Comp\Embi(k)\Comp\Embp(k)\in\Kl\PCOH(\Pnat,X)$ satisfies
\begin{align*}
  \Fun s(u)=\Fun t(u_0,\dots,u_{k-1},0,0,\dots)
\end{align*}
if we consider $\Pcoh\Snat$ as a subset of $\Realpto\Nat$.  We say that $t$
\emph{depends on at most $k$ parameters} if $t=t\Comp\Embi(k)\Comp\Embp(k)$,
which simply means that, for any
$(m,a)\in\Web{\Limpl{\Excl\Pnat}{X}}=\Mfin{\Nat}\times\Web X$, if
$t_{m,a}\not=0$ then $m\in\Mfin{\{0,\dots,k-1\}}$.

If $t\in\Kl\PCOH(\Pnat,X)$, $\Fun t$ is considered here as a function with
infinitely many real parameters. Given $k\in\Nat$ and $u\in\Realpto\Nat$, we
define $\Shvec uk\in\Realpto\Nat$ by $\Shvec uk_i=u_{i+k}$. Observe that
$s=t\Comp\Psem{\Shift_k}{}$ is characterized by $\Fun s(u)=\Fun t(\Shvec
uk)$. 

The term $\Shift_k$, as well as the other terms used below, is defined in
Section~\ref{sec:basic-examples}.

\paragraph{Testing term associated with a point of the web.}
\label{par:testing-contexts}
Given a type $\sigma$ and an element $a$ of $\Web{\Tsem\sigma}$, we define two
$\PCFP$ closed terms $\Ptest a$ and  $\Ntest a$ such that
\begin{align*}
  \Tseq{}{\Ptest a}{\Timpl\Tnat\sigma}
  \quad\text{and}\quad
  \Tseq{}{\Ntest a}{\Timpl\Tnat{\Timpl\sigma\Tnat}}\,.
\end{align*}
The definition is by mutual induction on $\sigma$. We first associate with $a$
two natural numbers $\Plen a$ and $\Nlen a$.

If $\sigma=\Tnat$, and hence $a=n\in\Nat$, we set
$\Plen a=\Nlen a=0$. 

If $\sigma=(\Timpl\phi\psi)$ so that $a=(\Mset{\List b1k},c)$ with
$b_i\in\Web{\Tsem\phi}$ for each $i=1,\dots,k$ and $c\in\Web{\Tsem\psi}$, we
set
\begin{align*}
  \Plen a &= \Plen c+\sum_{i=1}^k\Nlen{b_i}\\
  \Nlen a &= \Nlen c+k+\sum_{i=1}^k\Plen{b_i}
\end{align*}

Assume that $\sigma=\Tnat$, then $a=n$ for some $n\in\Nat$. We set
\begin{align*}
  \Ptest a=\Ptest n=\Abst\xi\Tnat{\Num n}\quad\text{and}\quad
  \Ntest a=\Ntest n=\Abst\xi\Tnat{\Probe_n}
\end{align*}
so that ${\Psem{\Ptest n}{}}(u)=n$ and 
${\Psem{\Ntest n}{}}(u)(w)=w_n\Snum 0$.

Assume that $\sigma=(\Timpl\phi\psi)$ so that $a=(\Mset{\List b1k},c)$ with
$b_i\in\Web{\Tsem\phi}$ for each $i=1,\dots,k$ and $c\in\Web{\Tsem\psi}$. Then
we define $\Ptest a$ such that $\Tseq{}{\Ptest a}{\Timpl\Tnat{\Timpl\phi\psi}}$
by
\begin{align*}
  \Ptest a=
  \Abstpref{\xi}{\Tnat}
  \Abstpref x\phi
  \IF(\ &
    \Apppref{\Pprod_k}\\
      &\hspace{2em}\Apppref{\Ntest{b_1}}\xi \Argsep x\\
      &\hspace{2em}\Apppref{\Ntest{b_2}}\Shvar{\Nlen{b_1}}{\xi}\Argsep x\\
      &\hspace{2em}\cdots\\
      &\hspace{2em}\Apppref{\Ntest{b_k}}\Shvar{\Nlen{b_1}+\cdots+\Nlen{b_{k-1}}}{\xi}\Argsep
      x,\\
    &\Apppref{\Ptest c}\Shvar{\Nlen{b_1}+\cdots+\Nlen{b_{k}}}{\xi},\\
    &[z]
    {\Omega_\psi}
    \ )
\end{align*}

Therefore we have, given $u\in\Pcoh\Pnat$ and $w\in\Pcoh{\Tsem\phi}$
\begin{align*}
  \Psem{\Ptest a}{}(u)(w)
  &= \Big(\prod_{i=1}^{k}\Psem{\Ntest{b_i}}{}
    \Big(\ShvecBig u{\sum_{j=1}^{i-1}\Nlen{b_j}}\Big)(w)\Big)_{\!0}
  \Psem{\Ptest c}{}\Big(\ShvecBig u{\sum_{j=1}^{k}\Nlen{b_j}}\Big)\,.
\end{align*}

The term $\Ntest a$ is such that
$\Tseq{}{\Ntest a}{\Timpl{\Tnat}{\Timpl{(\Timpl\phi\psi)}{\Tnat}}}$ and is
defined  by
\begin{align*}
  \Ntest a &=
  \Abstpref\xi\Tnat
  \Abstpref f{\Timpl\phi\psi}
  \Apppref{\Ntest c}
  \Shvar{k+\Plen{b_1}+\cdots+\Plen{b_{k}}}\xi\\
  &\hspace{2em}\Apppref f
  \Apppref{\Pchoose_k}\xi\Argsep\\
  &\hspace{6em}\Apppref{\Ptest{b_1}}{\Shvar k\xi}\\
  &\hspace{6em}\Argsep\cdots\Argsep\\
  &\hspace{6em}\Apppref{\Ptest{b_k}}
     {\Shvar{k+\Plen{b_1}+\cdots+\Plen{b_{k-1}}}\xi}\,.
\end{align*}
Therefore we have, given $u\in\Pcoh{\Pnat}$ and $t\in\Pcoh{(\Timpl\phi\psi)}$
\begin{align*}
  \Psem{\Ntest a}{}(u)(t)
  =\Psem{\Ntest c}{}(\ShvecBig u{k+\sum_{j=1}^k\Plen{b_j}})(\Fun t
  \Big(\sum_{i=1}^ku_{i-1}\Psem{\Ptest{b_i}}{}
    (\ShvecBig u{k+\sum_{j=1}^{i-1}\Plen{b_j}})
  \Big))\,.
\end{align*}

\begin{lemma}\label{lemma:test-depends}
  Let $\sigma$ be a type and $a\in\Web{\Tsem\sigma}$.
  Seen as an element of $\Kl\PCOH(\Pnat,\Tsem\sigma)$ (resp.~of
  $\Kl\PCOH(\Pnat,\Tsem{\Timpl\sigma\Tnat})$), $\Psem{\Ptest a}{}$
  (resp.~$\Psem{\Ntest a}{}$) depends on at most $\Plen a$ (resp.~$\Nlen a$)
  parameters. 
\end{lemma}
The proof is a simple induction on $\sigma$, based on an inspection of the
expressions above for $\Psem{\Ptest a}{}$ and $\Psem{\Ntest a}{}$.

\paragraph{More notations.}\label{par:more-notations}
Let $I=\{n_1<\cdots<n_k\}$ be a finite subset of $\Nat$, we use $\Msetu I$ for
the multiset $\Mset{\List n1k}$ where each element of $I$ appears exactly
once. Given $p,q\in\Nat$, we set
\begin{align*}
  \Msetb pq &= \Msetu{\{p,p+1,\dots,p+q-1\}}\\
  \Mfinr pq &= \Mfin{\{p,p+1,\dots,p+q-1\}}\,.
\end{align*}
These specific multisets, where each elements appears exactly once, play an
essential role in Lemma~\ref{lemma:FA-one-coef}.

Given $m\in\Mfin\Nat$ and $p$ and $q$ as above, we use the notation $\Restrms
mpq$ for the element $m'$ of $\Mfin\Nat$ defined by
\begin{align*}
  m'(i)=
  \begin{cases}
    m(i+p) & \text{if }0\leq i\leq q-1\\
    0 & \text{otherwise}
  \end{cases}
\end{align*}
and the notation $\Shleft mp$ for the element $m'$ of $\Mfin\Nat$ defined by
$m'(i)=m(i+p)$ for each $i\in\Nat$.

So $\Restrms mpq$ is obtained by selecting in $m$ a ``window'' starting at
index $p$ and ending at index $p+q-1$ and by shifting this window by $p$ to the
left. Similarly $\Shleft mp$ is obtained by shifting $m$ by $p$ to the left.

Given a set $I$ and an element $i$ of $I$, we use $\Canb i$ for the element of
$\Realpto I$ defined by $(\Canb i)_j=\Kronecker ij$.

\paragraph{Expression of the semantics of testing terms.}
\label{par:test-cont-expr}
We write now the functions $\Psem{\Ntest a}{}$ and $\Psem{\Ptest a}{}$ in a
form which makes explicit their dependency on their first argument
$u\in\Pcoh\Snat$. This also allows to make explicit their dependency on a
finite number of parameters.

Let $\sigma$ be a type and let $a\in\Web{\Tsem\sigma}$.  By
Lemma~\ref{lemma:test-depends}, for each $m\in\Mfinr 0{\Plen a}$, there are
uniquely defined $\pi(a,m)\in\Realpto{\Web{\Tsem\sigma}}$ and 
$\mu(a,m)\in\Realpto{\Web{(\Timpl{\Tsem\sigma}\Bot)}}$
such that we can write
  \begin{align}
    \Psem{\Ptest a}{}(u)=\sum_{m\in\Mfinr 0{\Plen a}}u^m\Prem am
    \label{eq:ptest-expr}
  \end{align}
for all $u\in\Pcoh\Snat$ and, for each $w\in\Pcoh{\Tsem\sigma}$,
  \begin{align}
    \Psem{\Ntest a}{}(u)(w)_0=\sum_{m\in\Mfinr 0{\Nlen
        a}}u^m\Nrem am(w) \,
    \label{eq:ntest-expr}
  \end{align}
for all $u\in\Pcoh\Snat$.

Observe that, for any $w\in\Pcoh{\Tsem\sigma}$, we have
\begin{align}
  \Nrem am(w)=\sum_{h\in\Mfin{\Web{\Tsem\sigma}}}\Nrem am_{(h,*)}w^h\,.
  \label{eq:nrem-expr}
\end{align}

\subsection{Proof of Full Abstraction}
We can now state and prove the main lemma in the proof of full
abstraction. This lemma uses notations introduced in
Section~\ref{sec:basic-notions}.

\begin{lemma}\label{lemma:FA-one-coef}
  Let $\sigma$ be a type and let $a\in\Web{\Tsem\sigma}$. We have
  \begin{align*}
    \Prem a{\Msetb 0{\Plen a}}&=\Canb a\\
    \Nrem a{\Msetb 0{\Nlen a}}&=\Canb{(\Mset a,*)}
  \end{align*}
  that is, $\Nrem a{\Msetb 0{\Nlen a}}(w)=w_a$ for each
  $w\in\Pcoh{\Tsem\sigma}$. 
\end{lemma}
\Beginproof
By induction on $\sigma$. Assume that $\sigma=\Tnat$ so that $a=n\in\Nat$ and
we have $\Plen n=\Nlen n=0$. We have $\Psem{\Ptest n}{}(u)=\Canb n$ and
$\Psem{\Ntest n}{}(u)(w)=w_n$ as expected.

Assume now that $\sigma=\Timpl\phi\psi$ so that $a$ can be written
\[
a=(\Mset{\List b1k},c)
\]
for some $\List b1k\in\Web{\Tsem\phi}$ and
$c\in\Web{\Tsem\psi}$. 

For each $u\in\Pcoh{\Snat}$ and
$w\in\Pcoh{\Tsem\phi}$, we have
\begin{align*}
  \Psem{\Ptest a}{}(u)(w)
  &=\prod_{i=1}^k\Bigg(\Psem{\Ntest{b_i}}{}
    (\ShvecBig u{\sum_{j=1}^{i-1}\Nlen{b_j}})(w)
      \Bigg)_{\!\!0}\Psem{\Ptest c}{}
    (\ShvecBig u{\sum_{j=1}^{k}\Nlen{b_j}})
    \quad\text{see \Parag{par:testing-contexts}}\\
  &=\prod_{i=1}^k\Bigg(
    \sum_{m\in\Mfinr{\sum_{j=1}^{i-1}\Nlen{b_j}}{\Nlen{b_i}}}
    u^m\Nrem{b_i}{\Shleft m{\sum_{j=1}^{i-1}\Nlen{b_j}}}(w)\Bigg)\\
  &\quad\quad\Bigg(\sum_{m\in\Mfinr{\sum_{j=1}^{k}
      \Nlen{b_j}}{\Plen c}}
  u^m\Prem c{\Shleft m{\sum_{i=1}^k\Nlen{b_i}}}\Bigg)
  \quad\text{see \Parag{par:test-cont-expr}}\\
  &=\sum_{m\in\Mfinr 0{\Plen a}}u^m
  \Big(\prod_{i=1}^k 
    \Nrem{b_i}{\Restrms m{\sum_{j=1}^{i-1}\Nlen{b_j}}{\Nlen{b_i}}}(w)\Big)\\
  &\hspace{16em}\Prem c{\Restrms m{\sum_{j=1}^{k}\Nlen{b_j}}{\Plen{c}}}\,,
\end{align*}
using the fact that $\Plen a=\sum_{j=1}^k\Nlen{b_i}+\Plen c$ and distributing
products over sums. We also use the fact that there is a bijection 
\begin{align*}
\Mfinr 0{\Plen a}
&\to
\Big(\prod_{i=1}^k\Mfinr{\sum_{j=1}^{i-1}\Nlen{b_j}}{\Nlen{b_i}}\Big)
\times\Mfinr{\sum_{j=1}^{k} \Nlen{b_j}}{\Plen c}  \\
m &\mapsto
((\Restrms m{\sum_{j=1}^{i-1}\Nlen{b_j}}{\Nlen{b_i}})_{i=1}^k,
    \Restrms m{\sum_{j=1}^{k}\Nlen{b_j}}{\Plen{c}})\,.
\end{align*}
Again we refer to \Parag{par:more-notations} for the notations used in these
expressions.

Therefore, given $m\in\Mfinr 0{\Plen a}$ and $w\in\Pcoh{\Tsem\phi}$, the
element $\Prem am(w)$ of 
$\Realp$ satisfies
\begin{align*}
  \Prem am(w)=\Big(\prod_{i=1}^k 
    \Nrem{b_i}{\Restrms m{\sum_{j=1}^{i-1}\Nlen{b_j}}{\Nlen{b_i}}}(w)\Big)
  \Prem c{\Restrms m{\sum_{j=1}^{k}\Nlen{b_j}}{\Plen{c}}}\,.
\end{align*}
In this expression, we take now $m=\Msetb 0{\Plen a}$. Since clearly $\Restrms
mpq=\Msetb 0q$ for all $p,q\in\Nat$ such that $p+q\leq\Plen a$, we get, by
inductive hypothesis:
\begin{align*}
  \Prem a{\Msetb 0{\Plen a}}(w)=\Big(\prod_{i=1}^k w_{b_i}\Big)\Canb c
\end{align*}
and hence $ \Prem a{\Msetb 0{\Plen a}}=\Canb a$ as contended.

Concerning $\Ntest a$, for each $u\in\Pcoh\Snat$ and
$t\in\Pcoh{\Tsem{\Timpl\phi\psi}}$, we have
\begin{align*}
  \Psem{\Ntest a}{}(u)(t)_0 &= \Psem{\Ntest c}{}(\ShvecBig
  u{k+\sum_{i=1}^k\Plen{b_i}})
  (\Fun t\Big(\sum_{i=1}^k u_{i-1}\Psem{\Ptest{b_i}}{}
  (\ShvecBig u{k+\sum_{j=1}^{i-1}\Plen{b_i}})\Big))_0\\
  &\hspace{24em}\text{see \Parag{par:testing-contexts}}\\
  &= \Psem{\Ntest c}{}(\ShvecBig u{k+\sum_{i=1}^k\Plen{b_i}})\\
  &\hspace{0.5em}(\Fun t\Big(\sum_{i=1}^ku_{i-1}
  \sum_{r\in\Mfinr{k+\sum_{j=1}^{i-1}
      \Plen{b_j}}{\Plen{b_i}}}u^r\Prem{b_i}{\Shleft{r}{k+\sum_{j=1}^{i-1}
      \Plen{b_j}}}\Big))_0\\
  &\hspace{24em}\text{see \Parag{par:test-cont-expr}}\\
 &=\sum_{\Biind{l\in\Mfinr{k+\sum_{i=1}^k\Plen{b_i}} {\Nlen
        c}}{h\in\Mfin{\Web{\Tsem\psi}}}}u^l
  \Nrem{c}{\Shleft{l}{k+\sum_{i=1}^k\Plen{b_i}}}_{(h,*)}\\
  &\hspace{2em}\Bigg(\sum_{(m',c')\in\Web{\Tsem{\Timpl\phi\psi}}}
  t_{m',c'}\Big(\sum_{i=1}^ku_{i-1}\\
  &\hspace{2em}\sum_{r\in\Mfinr{k+\sum_{j=1}^{i-1} \Plen{b_j}}{\Plen{b_i}}}
  u^r\Prem{b_i}{\Shleft r{k+\sum_{j=1}^{i-1}
      \Plen{b_j}}}\Big)^{m'}\Canb{c'}\Bigg)^h\\
  &\hspace{2em}\text{by \Parag{par:test-cont-expr}, \Eqref{eq:ntest-expr}, \Eqref{eq:nrem-expr} and
    by definition of
    application in $\Kl\PCOH$}\\
  &=\hspace{-3em}\sum_{\Biind{l\in\Mfinr{k+\sum_{i=1}^k\Plen{b_i}} {\Nlen
        c}}{h\in\Mfin{\Web{\Tsem\psi}}}}u^l
  \Nrem{c}{\Shleft{l}{k+\sum_{i=1}^k\Plen{b_i}}}_{(h,*)}
  \prod_{c'\in\Web{\Tsem\psi}} A(c')^{h(c')}
\end{align*}
where, for each $c'\in\Web{\Tsem{\psi}}$,
\begin{align*}
  A(c')&=\hspace{2em}
  \sum_{m'\in\Mfin{\Web{\Tsem\phi}}}
    t_{m',c'}\prod_{b\in\Web{\Tsem\phi}}\Big(\sum_{i=1}^ku^{\Mset{i-1}}\\
  &\hspace{2em}\sum_{r\in\Mfinr{k+\sum_{j=1}^{i-1}
        \Plen{b_j}}{\Plen{b_i}}}
    u^r\Prem{b_i}{\Shleft r{k+\sum_{j=1}^{i-1}
        \Plen{b_j}}}_b\Big)^{m'(b)}
\end{align*}
where we recall that $\Mset{i-1}$ is the multiset which has $i-1$ as unique
element. We can write $A(c')=\sum_{r\in\Mfin\Nat}u^r B(c')_r$ where $u$ does
not occur in the expression $B(c')_r$. For any $c'\in\Web{\Tsem{\psi}}$, all
the $r\in\Mfin\Nat$ such that $B(c')_r\not=0$ satisfy
$r\in\Mfinr{0}{k+\sum_{i=1}^k\Plen{b_i}}$: this results from a simple
inspection of the exponents of $u$ in the expression $A(c')$. It follows that,
for any $h\in\Mfin{\Web{\Tsem\psi}}$, we can write
\begin{align}
  \prod_{c'\in\Web{\Tsem\psi}}A(c')^{h(c')}
  =\sum_{r\in\Mfinr{0}{k+\sum_{i=1}^k\Plen{b_i}}}u^rD(r)_{h}
  \label{eq:expr-A}
\end{align}
where $u$ does not occur in the expressions $D(r)_h$. With these notations, we
have therefore
\begin{align*}
  \Psem{\Ntest a}{}(u)(t)_0
  &=\hspace{-2em}\sum_{\Biind{l\in\Mfinr{k+\sum_{i=1}^k\Plen{b_i}}
      {\Nlen c}}{h\in\Mfin{\Web{\Tsem\psi}}}}u^l
  \Nrem{c}{\Shleft{l}{k+\sum_{i=1}^k\Plen{b_i}}}_{(h,*)}
  \prod_{c'\in\Web{\Tsem\psi}}
  A(c')^{h(c')}\\
  &=\hspace{-1.5em}
  \sum_{\Biind{m\in\Mfinr{0}{\Nlen a}}{h\in\Mfin{\Web{\Tsem\psi}}}}
  u^m\Nrem{c}{\Shleft{m}{k+\sum_{i=1}^k\Plen{b_i}}}_h
  D(\Restrms{m}{0}{k+\sum_{i=1}^k\Plen{b_i}})_h
\end{align*}
In the second line, the $u^m$ results from the product
of the $u^l$ of the first line with the $u^r$ arising from~\Eqref{eq:expr-A}.
Remember indeed that $\Nlen a=k+\sum_{i=1}^k\Plen{b_i}+\Nlen c$.  We are
interested in the coefficient
\begin{align}
  \alpha=\Nrem{a}{\Msetb{0}{\Nlen a}}(t)
  \label{eq:alpha-def}
\end{align}
of $u^{\Msetb 0{\Nlen a}}$ in the sum above.
We have
\begin{align*}
  \alpha&=\sum_{h\in\Mfin{\Web{\Tsem\psi}}}
  \Nrem{c}{\Shleft{\Msetb{0}{\Nlen a}}{k+\sum_{i=1}^k\Plen{b_i}}}_h\\
  &\hspace{14em}D(\Restrms{\Msetb{0}{\Nlen a}}{0}{k+\sum_{i=1}^k\Plen{b_i}})_h\,.
\end{align*}
But $\Shleft{\Msetb{0}{\Nlen a}}{k+\sum_{i=1}^k\Plen{b_i}}=\Msetb{0}{\Nlen c}$
and hence, applying the inductive hypothesis to $c$, we get
\begin{align*}
  \alpha=D(\Msetb{0}{k+\sum_{i=1}^k\Plen{b_i}})_{\Mset c}\,.
\end{align*}
Coming back to~\Eqref{eq:expr-A}, we see that $\alpha$ is the coefficient of
$u^{\Msetb{0}{k+\sum_{i=1}^k\Plen{b_i}}}$ in $A(c)$ (indeed, in the present
situation $h=\Mset{c}$ and so the product which appears on the left side
of~\Eqref{eq:expr-A} has only one factor, namely $A(c)$).



So we focus our attention on $A(c)$, remember that
\begin{align*}
    A(c)&=\hspace{2em}
  \sum_{m'\in\Mfin{\Web{\Tsem\phi}}}
    t_{m',c}\prod_{b\in\Web{\Tsem\phi}}\Big(\sum_{i=1}^ku^{\Mset{i-1}}\\
  &\hspace{2em}\sum_{r\in\Mfinr{k+\sum_{j=1}^{i-1}
        \Plen{b_j}}{\Plen{b_i}}}
    u^r\Prem{b_i}{\Shleft r{k+\sum_{j=1}^{i-1}
        \Plen{b_j}}}_b\Big)^{m'(b)}\,.
\end{align*}

Let
\begin{align*}
  J=\Big\{(i,r)\St i\in\{1,\dots,k\}\text{ and }r\in\Mfinr{k+\sum_{j=1}^{i-1}
        \Plen{b_j}}{\Plen{b_i}}\Big\}\,.
\end{align*}
Observe that, given $(i,r),(i',r')\in J$, either $(i,r)=(i',r')$, or $i\not=i'$
and $r$ and $r'$ have disjoint supports.

Given $(i,r)\in J$, we set 
\begin{align}
  \theta(i,r)=\Prem{b_i}{\Shleft r{k+\sum_{j=1}^{i-1}\Plen{b_j}}}
  \label{eq:theta-def}
\end{align}
so that $\theta(i,r)\in\Realpto{\Web{\Tsem\phi}}$ for each $(i,r)\in J$.  With
these notations, we have
\begin{align*}
  A(c)&=\sum_{m'\in\Mfin{\Web{\Tsem\phi}}}
  t_{m',c}\prod_{b\in\Web{\Tsem\phi}}\Bigg(\sum_{(i,r)\in J}
  u^{\Mset{i-1}+r}\theta(i,r)_b\Bigg)^{m'(b)}\\
  &=\sum_{m'\in\Mfin{\Web{\Tsem\phi}}}
  t_{m',c}\prod_{b\in\Web{\Tsem\phi}}\Bigg(
  \sum_{\Biind{p\in\Mfin{J}}{\Card p=m'(b)}}
  u^{\sigma(p)}\Multinom{}{p}\theta_b^p\Bigg)
\end{align*}
where we recall that $\Multinom {}p=\Factor{(\Card
  p)}/\prod_{b\in\Web{\Tsem\phi}}\Factor{p(b)}$ is the multinomial coefficient
associated with the finite multiset $p$ by the multinomial formula. In this
expression, for each $b\in\Web{\Tsem\phi}$, $\theta_b$ is the $J$-indexed
family of real numbers defined by $\theta_b(i,r)=\theta(i,r)_b$ and
$\sigma(p)\in\Mfin\Nat$ is defined as
\begin{align}
  \sigma(p)=\sum_{(i,r)\in J}p(i,r)\cdot(\Mset{i-1}+r)\,.
  \label{eq:sigma-def}
\end{align}
Distributing the product over the sum and rearranging the sums, we get
\begin{align*}
  A(c)
  &=\sum_{m'\in\Mfin{\Web{\Tsem\phi}}}t_{m',c}
  \sum_{
    \Biind{\rho\in\Mfin{J}^{\Web{\Tsem\phi}}}
          {\forall b\ \Card{\rho(b)=m'(b)}}}
        u^{\sum_{b\in\Web{\Tsem\phi}}\sigma(\rho(b))}
        \prod_{b\in\Web{\Tsem\phi}}\Multinom{}{\rho(b)}\theta_b^{\rho(b)}\\
  &=\sum_{m\in\Mfinr{0}{k+\sum_{i=1}^{k}\Plen{b_i}}}
  u^m\sum_{\Biind{\rho\in\Mfin J^{\Web{\Tsem\phi}}}
    {\sum_{b\in\Web{\Tsem\phi}}\sigma(\rho(b))=m}}
  t_{\rho_1,c}\prod_{b\in\Web{\Tsem\phi}}\Multinom{}{\rho(b)}\theta_b^{\rho(b)}
\end{align*}
where
$\rho_1\in\Mfin{\Web{\Tsem\phi}}$ is defined by
\begin{align}
  \rho_1(b)=\Card{\rho(b)}=\sum_{(i,r)\in J}\rho(b)(i,r)
  \label{eq:rho1-def}
\end{align}
for each $\rho\in\Mfin J^{\Web{\Tsem\phi}}$%
. 
%
%
%
For
$m\in\Mfinr{0}{k+\sum_{i=1}^{k}\Plen{b_i}}$, let
\begin{align}
  \zeta(m)=\sum_{\Biind{\rho\in\Mfin J^{\Web{\Tsem\phi}}}
    {\sum_{b\in\Web{\Tsem\phi}}\sigma(\rho(b))=m}}
  t_{\rho_1,c}\prod_{b\in\Web{\Tsem\phi}}\Multinom{}{\rho(b)}\theta_b^{\rho(b)}
  \label{eq:zeta-def}
\end{align}
be the coefficient of $u^m$ in $A(c)$. 

Since we want to compute $\alpha=\zeta(\Msetb{0}{k+\sum_{i=1}^k\Plen{b_i}})$
defined in~\Eqref{eq:alpha-def}, we consider the particular case where
$m=\Msetb{0}{k+\sum_{i=1}^k\Plen{b_i}}$.  The elements $\rho$ of $\Mfin
J^{\Web{\Tsem\phi}}$ which index the sum~\Eqref{eq:zeta-def} satisfy the
condition $\sum_{b\in\Web{\Tsem\phi}}\sigma(\rho(b)) =\Msetb
0{k+\sum_{i=1}^k\Plen{b_i}}$, that is, coming back to the
definition~\Eqref{eq:sigma-def} of $\sigma$,
\begin{align}
  \sum_{\Biind{(i,r)\in J}{b\in\Web{\Tsem\phi}}}\rho(b)(i,r)\cdot(\Mset{i-1}+r)
  =\Msetb 0{k+\sum_{i=1}^k\Plen{b_i}}\,.
  \label{eq:cond-zeta}
\end{align}
Since $\Msetb
0{k+\sum_{i=1}^k\Plen{b_i}}=\Mset{0,\dots,k+\sum_{i=1}^k\Plen{b_i}-1}$
(see \Parag{par:more-notations}), condition~\Eqref{eq:cond-zeta} implies that,
for each $i\in\{1,\dots,k\}$, there is exactly one
$b_\rho(i)\in\Web{\Tsem\phi}$ and exactly one
$r_\rho(i)\in\Mfinr{k+\sum_{j=1}^{i-1}\Plen{b_j}}{\Plen{b_i}}$ such that
\[
\rho(b_\rho(i))(i,r_\rho(i))\not=0\,,
\]
and we know moreover that $\rho(b_\rho(i))(i,r_\rho(i))=1$ because $i-1$ occurs
exactly once in $\Mset{i-1}+r_\rho(i)$ (since the multisets $\Mset{i-1}$ and
$r_\rho(i)$ have disjoint supports for $i=1,\dots,k$). Moreover, since
$r_\rho(i)$ and $r_\rho(i')$ have disjoint supports when $i$ and $i'$ are
distinct elements of $\{1,\dots,k\}$, we must have
\begin{align}
  r_\rho(i)=\Msetb{k+\sum_{j=1}^{i-1}\Plen{b_j}}{\Plen{b_i}}
  \label{eq:mset-simple}
\end{align}
by \Eqref{eq:cond-zeta} again. 

From the first part of these considerations (existence and uniqueness of
$b_\rho(i)$ and $r_\rho(i)$), it follows that if $b\in\Web{\Tsem\phi}$ and
$(i,r)\in J$ are such that $\rho(b)(i,r)\not=0$ then we have $b=b_\rho(i)$ and
$r=r_\rho(i)$, and hence $\rho(b)(i,r)=1$. In particular,
$\Multinom{}{\rho(b)}=1$ for each $b$. It follows that
\begin{align*}
\prod_{b\in\Web{\Tsem\phi}}\Multinom{}{\rho(b)}\theta_b^{\rho(b)}
&=\prod_{b\in\Web{\Tsem\phi}}
\prod_{\Biind{(i,r)\in J}{b_\rho(i)=b,\ r_\rho(i)=r}}\theta(i,r)_b\\
&=\prod_{i=1}^k
  \Prem{b_i}{\Shleft{r_\rho(i)}{k+\sum_{j=1}^{i-1}\Plen{b_j}}}_{b_\rho(i)}  
\end{align*}
coming back to the definition of $\theta$, see~\Eqref{eq:theta-def}.
Let $H$ be the set of all $\rho$'s satisfying
\Eqref{eq:cond-zeta}, we have therefore
\begin{align*}
  \alpha=\zeta(\Msetb 0{k+\sum_{i=1}^k\Plen{b_i}})
  &=\sum_{\rho\in H}
  t_{\rho_1,c}\prod_{b\in\Web{\Tsem\phi}}
    \Multinom{}{\rho(b)}\theta_b^{\rho(b)}\\
  &=\sum_{\rho\in H}
  t_{\rho_1,c}\prod_{i=1}^k
  \Prem{b_i}{\Shleft{r_\rho(i)}{k+\sum_{j=1}^{i-1}\Plen{b_j}}}_{b_\rho(i)}\\
  &\hspace{8em}\text{by the observations above}\\
  &=\sum_{\rho\in H}
  t_{\rho_1,c}\prod_{i=1}^k
  \Prem{b_i}{\Msetb 0{\Plen{b_i}}}_{b_\rho(i)}
  \quad\text{by~\Eqref{eq:mset-simple}.}
\end{align*}
By our inductive hypothesis about $\Prem{b_i}{\Msetb 0{\Plen{b_i}}}$, all the
terms of this sum vanish, but the one corresponding to the unique element
$\rho$ of $H$ such that $b_\rho(i)=b_i$ for $i=1,\dots,k$. For this specific
$\rho$, coming back to the definition~\Eqref{eq:rho1-def} of $\rho_1$, we have
$\rho_1=\Mset{\List b1k}$. It follows that
\begin{align*}
  \alpha=\zeta(\Msetb 0{k+\sum_{i=1}^k\Plen{b_i}})=t_{\Mset{\List b1k},c}
\end{align*}
as contended, and this ends the proof of the lemma.
\Endproof

\paragraph{Main statements.} 
We first state a separation theorem which seems interesting on its own right
and expresses that our testing terms $\Ntest a$, when fed with suitable
rational probability distributions, are able to separate any two distinct
elements of the interpretation of a type.
\begin{theorem}[Separation]\label{th:FA-test-separation}
  Let $\sigma$ be a type and let $a\in\Web{\Tsem\sigma}$. Let
  $w,w'\in\Pcoh{\Tsem\sigma}$ be such that $w_a\not=w'_a$. Let $n=\Nlen
  a$. There is a sequence $(q_i)_{i=0}^{n-1}$ of rational numbers such that the
  element $u=\sum_{i=0}^{n-1}q_i\Canb i$ of $\Pcoh{\Snat}$ satisfies
  $\Psem{\Ntest a}{}(u)(w)\not=\Psem{\Ntest a}{}(u)(w')$.
\end{theorem}
\Beginproof
With the notations of the statement of the proposition, we consider the
functions $\phi,\phi':\Pcoh{\Snat}\to \Realp$ defined by $\phi(u)=\Psem{\Ntest
  a}{}(u)(w)_0$ and $\phi'(u)=\Psem{\Ntest a}{}(u)(w')_0$. By
Lemma~\ref{lemma:test-depends}, the morphisms $\phi$ and $\phi'$ depend on at
most $n=\Nlen a$ parameters. In other words, there are $t,t'\in\Kl\PCOH(\Simplex
n,\Bot)$ such that
\begin{align*}
  \forall u\in\Pcoh{\Snat}\quad
  \phi(u)=\Fun t\Big(\sum_{i=0}^{n-1}u_i\Canb i\Big)
  \text{ and }
  \phi'(u)=\Fun{t'}\Big(\sum_{i=0}^{n-1}u_i\Canb i\Big)
\end{align*}
Coming back to~\Eqref{eq:ntest-expr}, we see that the coefficient of
$u^{\Mset{0,\dots,n-1}}$ in the expression of $\phi(u)$ is $\Nrem
a{\Mset{0,\dots,n-1}}(w)$, whose value is $w_a$ by
Lemma~\ref{lemma:FA-one-coef}. In other words $t_{\Mset{0,\dots,n-1},*}=w_a$
and similarly $t'_{\Mset{0,\dots,n-1},*}=w'_a$. From this, it results that the
functions $\Fun t$ and $\Fun{t'}$ from $\Pcoh{\Simplex n}$ to $\Real$ are
distinct (because these are entire functions with distinct power series, which
are defined on the subset $\Pcoh{\Simplex n}$ of $\Realpto n$, which contains a
non-empty subset of $\Real^n$ which is open for the usual topology). Since
these functions are continuous (again, for the usual topology), there is an
$u\in\Pcoh{\Simplex n}$ such that $\List u0{n-1}\in\Rational$ and $\Fun
t(u)\not=\Fun{t'}(u)$.
\Endproof

\begin{theorem}[Full Abstraction]
  Let $\sigma$ be a type, $\Gamma$ be a typing context and let $M$ and $M'$ be
  terms such that $\Tseq{\Gamma}M\sigma$ and $\Tseq{\Gamma}{M'}\sigma$. If
  $M\Rel\Obseq M'$ then $\Psem M{\Gamma}=\Psem{M'}{\Gamma}$.
\end{theorem}
\Beginproof
Assume that $\Psem M\Gamma\not=\Psem{M'}\Gamma$.

Let $(x_1:\sigma_1,\dots,x_k:\sigma_k)$ be the typing context $\Gamma$.  Let
$N=\Abst{x_1}{\sigma_1}{\cdots\Abst{x_k}{\sigma_k}{M}}$ and
$N'=\Abst{x_1}{\sigma_1}{\cdots\Abst{x_k}{\sigma_k}{M'}}$ be closures of $M$
and $M'$. Let $\tau=\Timpl{\sigma_1}{\cdots\Timpl{\sigma_k}{\sigma}}$.

Let $w=\Psem N{}$ and $w'=\Psem{N'}{}$, we have $w\not=w'$ so there is
$a\in\Web{\Tsem\tau}$ such that $w_a\not=w'_a$. By
Theorem~\ref{th:FA-test-separation}, we can find a sequence $(q_i)_{i=0}^{n-1}$
of rational numbers such that for all $i\in\{0,\dots,n-1\}$ one has $q_i\geq 0$
and $\sum_{i=0}^{n-1}q_i\leq 1$, and $u=\sum_{i=0}^{n-1}q_i\Canb
i\in\Pcoh{\Snat}$ satisfies $\Psem{\Ntest a}{}(u)(w)_0\not=\Psem{\Ntest
  a}{}(u)(w')_0$.

Observe that $u=\Psem{\Ran{\List q0{n-1}}}{}$.

Let $C$ be the following observation context: 
\[
\Thole C\Gamma\sigma=\App{\Ntest a}{\Ran{\List q0{n-1}}\Argsep
\Abst{x_1}{\sigma_1}{\cdots\Abst{x_k}{\sigma_k}{\Hole{\Gamma}{\sigma}}}}
\]
which satisfies $\Tseqh{}{C}{\Gamma}{\sigma}{\Tnat}$, $\Psem{\Thsubst
  CM}{}=\Psem{\Ntest a}{}(u)(w)$ and $\Psem{\Thsubst C{M'}}{}=\Psem{\Ntest
  a}{}(u)(w')$.

Applying Theorem~\ref{th:Nat-sem-proba}, we get that
\begin{align*}
  \Redmat\Tnat^\infty_{\Thsubst CM,\Num 0}\not=\Redmat\Tnat^\infty_{\Thsubst
    C{M'},\Num 0}
\end{align*}
which shows that $M\Rel{\not\Obseq}M'$.
\Endproof

\paragraph{Failure of inequational full abstraction.}
We can define an observational preorder on closed terms: given terms $M$ and
$M'$ such that $\Tseq{}{M}{\sigma}$ and $\Tseq{}{M'}{\sigma}$, let us write
$M\Obsleq M'$ if, for all closed $C$ such that $\Tseq{}{C}{\Impl\sigma\Tnat}$,
one has $\Redmat{\Tnat}^\infty_{\App{C}{M},\Num
  0}\leq\Redmat{\Tnat}^\infty_{\App{C}{M'},\Num 0}$. Then it is easy to see
that $\Psem{M}{}\leq\Psem{M'}{}\Implies M\Obsleq M'$ (just as in the proof of
Theorem~\ref{th:eq-adequacy}).

The converse implication however is far from being true. A typical
counter-example (which is essentially the same as the example of the
Remark following Proposition~\ref{prop:order-fun-kleiseli}) is provided by the
two terms
\begin{align*}
  M_1 &= \Abst x\Tnat{\Ifv{x}{\Num 0}{z}{\Loopt\Tnat}}\\
  M_2 &= \Abst x\Tnat{\Ifv{x}{\Ifv{x}{\Num 0}{z'}{\Loopt\Tnat}}
    {z}{\Loopt\Tnat}}
\end{align*}
One has $\Tseq{}{M_i}{\Timpl{\Tnat}{\Tnat}}$ for $i=1,2$ and the functional
behavior of the interpretations of these terms is given by
\begin{align*}
  \Psem{M_1}{}(u) &= u_0\,\Snum 0\\
  \Psem{M_2}{}(u) &= u_0^2\,\Snum 0
\end{align*}
for all $u\in\Pcoh\Snat$ so $\Psem{M_1}{}$ and $\Psem{M_2}{}$ are not
comparable in $\Pcoh{\Tsem{\Timpl\Tnat\Tnat}}$ and nevertheless one can check
that $M_2\Obsleq M_1$. The proof boils down to the observation that, for each
$u\in\Pcoh\Snat$, one has $\Psem{M_2}{}(u)\leq\Psem{M_1}{}(u)$.

\section*{Conclusion}
We have studied an operationally meaningful probabilistic extension of PCF and,
in particular, we have proven a full abstraction result for the probabilistic
coherence spaces model of Linear Logic, with respect to a natural notion of
observational equivalence on the terms of this language.

This observational equivalence can be considered as too restrictive however
since it is based on a strict equality of probabilities of convergence.  In the
present probabilistic setting, a suitable \emph{distance} on terms could
certainly be more relevant, and provide more interesting information on the
behavior of programs, than our observational equivalence relation. The study of
such notions of distance and of their connections with PCSs, based on earlier
works by various authors, will be the purpose of our next investigations. We
also plan to extend our adequacy and, if possible, full abstraction results to
richer type structures, in a call-by-push-value flavored setting.


\bibliographystyle{alpha}
\bibliography{newbiblio}

\end{document}

%% file: local.tex
\newcommand\CMLLPAR{
\usepackage{cmll}
\newcommand\IPar{\mathord{\parr}}
}

\CMLLPAR

%% file: notation.tex
\usepackage{stmaryrd}

\newtheorem{theorem}{Theorem}

\newtheorem{proposition}[theorem]{Proposition}
\newtheorem{lemma}[theorem]{Lemma}

\theorembodyfont{\normalfont}
\theoremstyle{plain}

\makeatletter%
\renewcommand\subsubsection{%
\@startsection{subsubsection}{3}{0mm}{\baselineskip}{-1em}%
{\normalfont\bfseries}}
\makeatother
\renewcommand\paragraph{\subsubsection}

\newcommand\Proofbreak{\smallbreak}

\newenvironment{example}%
{\smallbreak\noindent{\bf Example.}\nobreak}%
{\normalsize\smallbreak}

\newcommand{\proofitem}[1]{\paragraph*{\mdseries\textit{#1}}}

\newcommand{\Beginproof}{\proofitem{Proof.}}
\newcommand{\Endproof}{
  \ifmmode 
  \else \leavevmode\unskip\penalty9999 \hbox{}\nobreak\hfill
  \fi
  \quad\hbox{$\Box$}
  \par\medskip}

\newcommand\Eqref[1]{(\ref{#1})}

\newenvironment{remark}%
{\smallbreak\noindent{\textit{Remark\/}: }\nobreak}%
{\smallbreak}

\renewcommand{\phi}{\varphi}
\renewcommand\epsilon{\varepsilon}

\newcommand{\Implies}{\Rightarrow}

\newcommand{\St}{\mid}

\renewcommand{\Bot}{{\mathord{\perp}}}
\newcommand{\Top}{\top}

\newcommand\cF{\mathcal{F}}

\newcommand\cR{\mathcal{R}}

\newcommand\cX{\mathcal{X}}
\newcommand\cY{\mathcal{Y}}

\newcommand\Fini{{\mathrm{fin}}}

\newcommand\Union{\bigcup}

\newcommand{\Linarrow}{\multimap}

\newcommand\Myleft{}
\newcommand\Myright{}

\newcommand\Web[1]{\Myleft|{#1}\Myright|}

\newcommand\Supp[1]{\operatorname{\mathsf{supp}}({#1})}

\newcommand\Mset[1]{[{#1}]}

\newcommand\ITens{\otimes}
\newcommand\Tens[2]{{#1}\ITens{#2}}
\newcommand\Tensp[2]{\left({#1}\ITens{#2}\right)}
\newcommand\IWith{\mathrel{\&}}
\newcommand\With[2]{{#1}\IWith{#2}}
\newcommand\IPlus{\oplus}
\newcommand\Plus[2]{{#1}\IPlus{#2}}
\newcommand\Orth[2][]{#2^{\Bot_{#1}}}
\newcommand\Orthp[2][]{(#2)^{\Bot_{#1}}}

\newcommand\Bwith{\mathop{\&}}
\newcommand\Bplus{\mathop\oplus}

\newcommand\Inj[1]{\overline\pi_{#1}}

\newcommand\Biorth[1]{#1^{\Bot\Bot}}

\newcommand\Triorth[1]{{#1}^{\Bot\Bot\Bot}}

\newcommand\One{1}

\newcommand\Restr[2]{{#1}|_{#2}}

\newcommand\Card[1]{\#{#1}}

\newcommand\Locun[1]{1^J}

\newcommand\Isom\simeq

\newcommand\Comp{\mathrel\circ}

\newcommand\Funinv[1]{{#1}^{-1}}

\newcommand\Limpl[2]{{#1}\Linarrow{#2}}

\newcommand\Nat{{\mathbb{N}}}

\newcommand\Natnz{{\Nat^+}}

\newcommand\Biind[2]{\genfrac{}{}{0pt}{1}{#1}{#2}}

\newcommand\Snat{\mathsf N}

\newcommand\App[2]{\left({#1}\right){#2}}
\newcommand\Apppref[1]{\left({#1}\right)}
\newcommand\Appp[3]{\left({#1}\right){#2}\,{#3}}

\newcommand\Abst[3]{\lambda#1^{#2}\,{#3}}

\newcommand\Parag[1]{\S\ref{#1}}

\newcommand\List[3]{#1_{#2},\dots,#1_{#3}}

\newcommand\Kronecker[2]{\delta_{{#1},{#2}}}

\newcommand\Subst[3]{{#1}\left[{#2}/{#3}\right]}

\newcommand\Substbis[2]{{#1}\left[{#2}\right]}

\newcommand\Factor[1]{{#1}!}

\newcommand\Multinom[2]{\mathsf{mn}(#2)}

\newcommand\Real{\mathbb{R}}
\newcommand\Realp{\mathbb{R}^+}
\newcommand\Realpto[1]{(\Realp)^{#1}}

\newcommand\Realpc{\overline{\mathbb{R}^+}}
\newcommand\Realpcto[1]{\Realpc^{#1}}

\newcommand\Rational{\mathbb Q}

\newcommand\Mfin[1]{\mathcal M_\Fini({#1})}
\newcommand\Mfinc[2]{\mathcal M_{#1}({#2})}

\newcommand\Ev{\operatorname{\mathsf{Ev}}}
\newcommand\Evlin{\operatorname{\mathsf{ev}}}

\newcommand\Norm[1]{\|{#1}\|}

\newcommand\Rel[1]{\mathrel{#1}}

\newcommand\Redst[1]{\mathop{\mathsf{Red}}}

\newcommand\Tuple[1]{\langle{#1}\rangle}

\newcommand\Msetofsubst[1]{\bar F}

\newcommand\Pcoh[1]{\mathsf P{#1}}
\newcommand\Pcohp[1]{\mathsf P(#1)}

\newcommand\Matapp[2]{{#1}\Compl{#2}}

\newcommand\PCOH{\mathbf{Pcoh}}

\newcommand\Leftu{\lambda}
\newcommand\Rightu{\rho}
\newcommand\Assoc{\alpha}
\newcommand\Sym{\gamma}
\newcommand\Msetempty{\Mset{\,}}

\newcommand\Retri\zeta
\newcommand\Retrp\rho

\newcommand\Impl[2]{{#1}\Rightarrow{#2}}

\newcommand\Tsem[1]{\llbracket{#1}\rrbracket}
\newcommand\Tseme[1]{\llbracket{#1}\rrbracket^{\mathord\oc}}
\newcommand\Psem[2]{\llbracket{#1}\rrbracket_{#2}}

\newcommand\Tnat\iota
\newcommand\Fix[1]{\operatorname{\mathsf{fix}}(#1)}
\newcommand\If[3]{\operatorname{\mathsf{if}}(#1,#2,#3)}
\newcommand\Pred{\operatorname{\mathsf{pred}}}
\newcommand\Succ[1]{\operatorname{\mathsf{succ}}(#1)}
\newcommand\Num[1]{\underline{#1}}
\newcommand\Loop\Omega
\newcommand\Loopt[1]{\Omega^{#1}}
\newcommand\Ran[1]{\mathsf{ran}(#1)}
\newcommand\Dice[1]{\operatorname{\mathsf{coin}}(#1)}
\newcommand\Tseq[3]{{#1}\vdash{#2}:{#3}}

\newcommand\Timpl\Impl
\newcommand\Simpl\Impl

\newcommand\PCFP{\mathsf{pPCF}}
\newcommand\PCF{\mathsf{PCF}}
\newcommand\PCFPZ{\mathsf{pPCF}^-}

\newcommand\Pnat{\mathbf N}

\newcommand\Redone[1]{\stackrel{#1}\rightarrow}

\newcommand\Redoned{\rightarrow_{\mathsf d}}

\newcommand\Weak[1]{\operatorname{\mathsf{w}}_{#1}}

\newcommand\Contrc[1]{\operatorname{\mathsf{contr}}_{#1}}

\newcommand\Der[1]{\operatorname{\mathsf{der}}_{#1}}

\newcommand\Digg[1]{\operatorname{\mathsf{dig}}_{#1}}

\newcommand\Fun[1]{\widehat{#1}}

\newcommand\Id{\operatorname{\mathsf{Id}}}

\newcommand\Proj[1]{\pi_{#1}}

\newcommand\Excl[1]{\oc{#1}}

\newcommand\Prom[1]{#1^!}

\newcommand\Relincl\eta
\newcommand\Relrestr\rho

\newcommand\Seely{\mu}

\newcommand\Monoidal{\mathsf m}

\newcommand\Compl{\,}
\newcommand\Curlin{\mathsf{cur}}
\newcommand\Cur{\mathsf{Cur}}

\newcommand\Kl[1]{{#1}_\oc}

\newcommand\IF{\mathsf{if}}

\newcommand\Ifv[4]{\IF(#1,#2,#3\cdot #4)}

\newcommand\Abstpref[2]{\lambda #1^{#2}\,}

\newcommand\Convproba[1]{\mathrel{\downarrow^{#1}}}

\newcommand\Transcl[1]{{#1}^*}

\newcommand\Eval[2]{\langle#1,#2\rangle}

\newcommand\Let[3]{\mathsf{let}\ #1\ \mathsf{be}\ #2\ \mathsf{in}\ #3}

\newcommand\Add{\mathsf{add}}
\newcommand\Exp{\mathsf{exp}_2}
\newcommand\Cmp{\mathsf{cmp}}

\newcommand\Unift{\mathsf{unif}_2}
\newcommand\Unif{\mathsf{unif}}

\newcommand\Closed[1]{\Lambda^{#1}_0}
\newcommand\Open[2]{\Lambda^{#2}_{#1}}

\newcommand\Redmat[1]{\mathsf{Red}(#1)}
\newcommand\Redmats{\mathsf{Red}}
\newcommand\Redmato[2]{\mathsf{Red}(#1,#2)}

\newcommand\Mexpset[2]{\mathsf L(#1,#2)}

\newcommand\Expmonisoz{\Seely^0}
\newcommand\Expmonisob[2]{\Seely^2_{#1,#2}}
\newcommand\Expmonisobn{\Seely^2}

\newcommand\Injms[2]{#1\cdot#2}

\newcommand\Natobj{\mathsf{N}}
\newcommand\Natalg{h_\Natobj}

\newcommand\Snum[1]{\overline{#1}}
\newcommand\Sif{\overline{\mathsf{if}}}
\newcommand\Ssuc{\overline{\mathsf{suc}}}

\newcommand\Sfix{\mathsf{Y}}

\newcommand\Vect[1]{\vec{#1}}

\newcommand\Rts[1]{\cR^{#1}}

\newcommand\Probw[1]{\mathsf p(#1)}
\newcommand\Spath[2]{\mathsf{R}(#1,#2)}

\newcommand\Obseq{\sim}
\newcommand\Obsleq{\lesssim}

\newcommand\Ptest[1]{{#1}^+}
\newcommand\Ntest[1]{{#1}^-}
\newcommand\Plen[1]{|#1|^+}
\newcommand\Nlen[1]{|#1|^-}

\newcommand\Shift{\mathsf{shift}}
\newcommand\Shvar[2]{\App{\Shift_{#1}}{#2}}

\newcommand\Probe{\mathsf{prob}}
\newcommand\Pprod{\mathsf{prod}}
\newcommand\Pchoose{\mathsf{choose}}

\newcommand\Argsep{\,}

\newcommand\Simplex[1]{\Delta_{#1}}

\newcommand\Embi{\eta^+}
\newcommand\Embp{\eta^-}

\newcommand\Shvec[2]{{#1}\left\{{#2}\right\}}
\newcommand\ShvecBig[2]{{#1}\Big\{{#2}\Big\}}

\newcommand\Msetu[1]{\mathsf o(#1)}
\newcommand\Msetb[2]{\mathsf o(#1,#2)}

\newcommand\Mfinr[2]{\Mfin{#1,#2}}

\newcommand\Canb[1]{\mathsf e_{#1}}

\newcommand\Prem[2]{\pi(#1,#2)}
\newcommand\Nrem[2]{\mu(#1,#2)}

\newcommand\Restrms[3]{\mathsf W(#1,#2,#3)}
\newcommand\Shleft[2]{\mathsf S(#1,#2)}

\newcommand\Listarg[3]{\,#1_{#2}\cdots#1_{#3}}

\newcommand\Rseg[2]{[#1,#2]}

\newcommand\Iftrans[2]{#1^\bullet_{#2}}

\newcommand\Bnfeq{\mathrel{\mathord:\mathord=}}
\newcommand\Bnfor{\,\,\mathord|\,\,}

\newcommand\PPCF{\mathsf{pPCF}}

\newcommand\Hempty{\ \ }
\newcommand\Hole[2]{[\Hempty]^{#1\vdash #2}}
\newcommand\Thole[3]{#1^{#2\vdash #3}}
\newcommand\Thsubst[2]{#1[#2]}

\newcommand\Tseqh[5]{\Tseq{#1}{\Thole{#2}{#3}{#4}}{#5}}

%% file: main.bbl
\newcommand{\SortNoop}[1]{}
\begin{thebibliography}{Ehr15b}

\bibitem[Bie95]{Bierman95}
Gavin Bierman.
\newblock What is a categorical model of intuitionistic linear logic?
\newblock In Mariangiola Dezani-Ciancaglini and Gordon~D. Plotkin, editors,
  {\em Proceedings of the second Typed Lambda-Calculi and Applications
  conference}, volume 902 of {\em {Lecture Notes in Computer Science}}, pages
  73--93. Springer-Verlag, 1995.

\bibitem[DE11]{DanosEhrhard08}
Vincent Danos and Thomas Ehrhard.
\newblock {Probabilistic coherence spaces as a model of higher-order
  probabilistic computation}.
\newblock {\em {Information and Computation}}, 152(1):111--137, 2011.

\bibitem[Ehr02]{Ehrhard00c}
Thomas Ehrhard.
\newblock On {K\"othe} sequence spaces and linear logic.
\newblock {\em {Mathematical Structures in Computer Science}}, 12:579--623,
  2002.

\bibitem[Ehr15a]{Ehrhard15b}
Thomas Ehrhard.
\newblock {A Call-By-Push-Value FPC and its interpretation in Linear Logic}.
\newblock Technical report, University Paris Diderot, PPS Laboratory, 2015.

\bibitem[Ehr15b]{Ehrhard15c}
Thomas Ehrhard.
\newblock {A semantical introduction to differential linear logic}.
\newblock {\em {Mathematical Structures in Computer Science}}, 2015.
\newblock To appear.

\bibitem[EPT11]{EhrhardPaganiTasson11}
Thomas Ehrhard, Michele Pagani, and Christine Tasson.
\newblock {The computational meaning of probabilistic coherent spaces}.
\newblock In {\em {Proceedings of the 26th Annual IEEE Symposium on Logic in
  Computer Science, LICS 2011, June 21-24, 2011, Toronto, Ontario, Canada}},
  pages 87--96. {IEEE Computer Society}, 2011.

\bibitem[ETP14]{EhrhardPaganiTasson14}
Thomas Ehrhard, Christine Tasson, and Michele Pagani.
\newblock {Probabilistic coherence spaces are fully abstract for probabilistic
  PCF}.
\newblock In Suresh Jagannathan and Peter Sewell, editors, {\em POPL}, pages
  309--320. ACM, 2014.

\bibitem[Gir88]{Girard88c}
Jean-Yves Girard.
\newblock Normal functors, power series and the $\lambda$-calculus.
\newblock {\em {Annals of Pure and Applied Logic}}, 37:129--177, 1988.

\bibitem[Gir99]{Girard99}
Jean-Yves Girard.
\newblock Coherent {B}anach spaces: a continuous denotational semantics.
\newblock {\em {Theoretical Computer Science}}, 227(1-2):275--297, 1999.

\bibitem[Gir04]{Girard04a}
Jean-Yves Girard.
\newblock Between logic and quantic: a tract.
\newblock In Thomas Ehrhard, Jean-Yves Girard, Paul Ruet, and Philip Scott,
  editors, {\em Linear Logic in Computer Science}, volume 316 of {\em {London
  Mathematical Society Lecture Notes Series}}, pages 346--381. {Cambridge
  University Press}, 2004.

\bibitem[JT98]{JungTix98}
Achim Jung and Regina Tix.
\newblock The troublesome probabilistic powerdomain.
\newblock {\em {Electronic Notes in Theoretical Computer Science}}, 13:70--91,
  1998.

\bibitem[Lev06]{LevyP06}
Paul~Blain Levy.
\newblock Call-by-push-value: Decomposing call-by-value and call-by-name.
\newblock {\em {Higher-Order and Symbolic Computation}}, 19(4):377--414, 2006.

\bibitem[Mel09]{Mellies09}
Paul-Andr\'e Melli\`es.
\newblock {Categorical semantics of linear logic}.
\newblock {\em Panoramas et Synth\`eses}, 27, 2009.

\bibitem[Plo77]{Plotkin77}
Gordon Plotkin.
\newblock {LCF} considered as a programming language.
\newblock {\em {Theoretical Computer Science}}, 5:223--256, 1977.

\end{thebibliography}
